\documentclass[12pt]{article}
\usepackage{graphicx}
\usepackage{subfigure}
\usepackage{cite}

\newcommand\pubnumber{SNSN-323-63}
\newcommand\pubdate{\today}

\def\ihep{Institute of High Energy Physics, Beijing, China}

\def\Title#1{\begin{center} {\Large #1 } \end{center}}
\def\Author#1{\begin{center}{ \sc #1} \end{center}}
\def\Address#1{\begin{center}{ \it #1} \end{center}}

\newcommand\pubblock{\rightline{\begin{tabular}{l} \pubnumber\\
         \pubdate  \end{tabular}}}
\newenvironment{Abstract}{\begin{quotation}  }{\end{quotation}}
\newenvironment{Presented}{\begin{quotation} \begin{center} 
             PRESENTED AT\end{center}\bigskip 
      \begin{center}\begin{large}}{\end{large}\end{center} \end{quotation}}
\def\Acknowledgements{\bigskip  \bigskip \begin{center} \begin{large}
             \bf ACKNOWLEDGEMENTS \end{large}\end{center}}
\def\qqbar{$q\bar{q}$}
\def\qqbarq{$q\bar{q}q$}
\def\ppbar{$p\bar{p}$}
\def\xppbar{$X(p\bar{p})$}
\def\jpsi{$J/\psi$}

\def\etal{{\it et al.}}
\def\ie{{\it i.e.}}

\begin{document}
\begin{titlepage}
\pubblock

\vfill
\Title{Exotic in Leptonic Machines}
\vfill
\Author{Kai Zhu}
\Address{\ihep}
\vfill
\begin{Abstract}
Selected topics of exotics in leptonic machines are presented, including recent discovery
of abnormal structures around the $p\bar{p}$ threshold and new information of the XYZ
(charmonium-like) states.
\end{Abstract}
\vfill
\begin{Presented}
Flavor Physics \& CP Violation 2014 (FPCP 2014)\\
Marseille, France, May 25--30, 2014
\end{Presented}
\vfill
\end{titlepage}
\def\thefootnote{\fnsymbol{footnote}}
\setcounter{footnote}{0}

\section{Introduction}
After Higgs, the last missing block of the standard model (SM), has been discovered
experimentally, more physicists transfer their interest to new physics beyond the
SM. However, even within the SM model, there are still some predictions have not been
confirmed or validated. The quantum chromodynamics (QCD) predicts that there are more
hadronic states than the usual mesons (\qqbar~states) and baryons ($qqq$~states). There
should be bound gluons (glueball), \qqbar-pair with an excited gluon (hybrids),
multi-quark color singlet states such as: \qqbar\qqbar~(tetra-quark and molecular),
\qqbar\qqbarq~(penta-quark), \qqbar\qqbar\qqbar~(six-quark and baryonium) and
\etal. However, none of these exotic states have really been established or ruled out
experimentally.

Leptonic machines, so called charm-factories or B-factories, due to their relative high
luminosities and clean backgrounds, provide idea places to study these exotic states. In
this proceeding I will present some new the discoveries and measurements attributed to
BESIII, CLEOc, Belle and BarBar. The main body will be divided into two parts, one is for
the light-hadronic spectrum and the other one is for heavier charmonium-likes states. The
first part contains the recently found structures near \ppbar~threshold such as \xppbar,
$X(1835)$, $X(1840)$, $X(1870)$, $X(1810)$. The second part contains the very recent
progresses on the XYZ (charmonium-like) states, such as new production mode, discoveries of
new resonance, updating previous measurements with larger statistics, \etal. I must
apologize I cannot include all the interesting topics in this field.

\section{Structures near the \ppbar~ threshold}
\begin{figure}[ht!]
\centering
\subfigure[]{\label{fig:1a}\includegraphics[width=0.3\textwidth,height=0.27\textwidth]{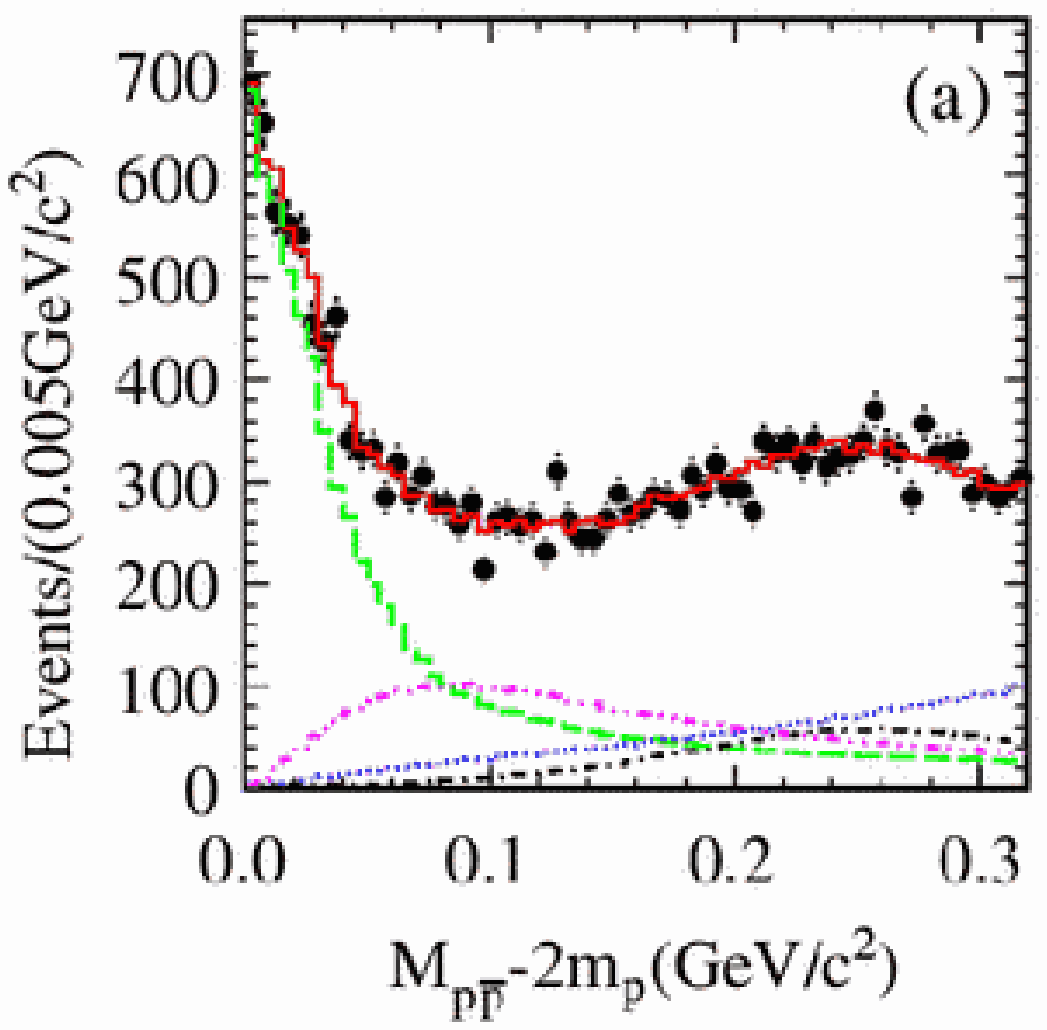}}
\subfigure[]{\label{fig:1b}\includegraphics[width=0.3\textwidth,height=0.27\textwidth]{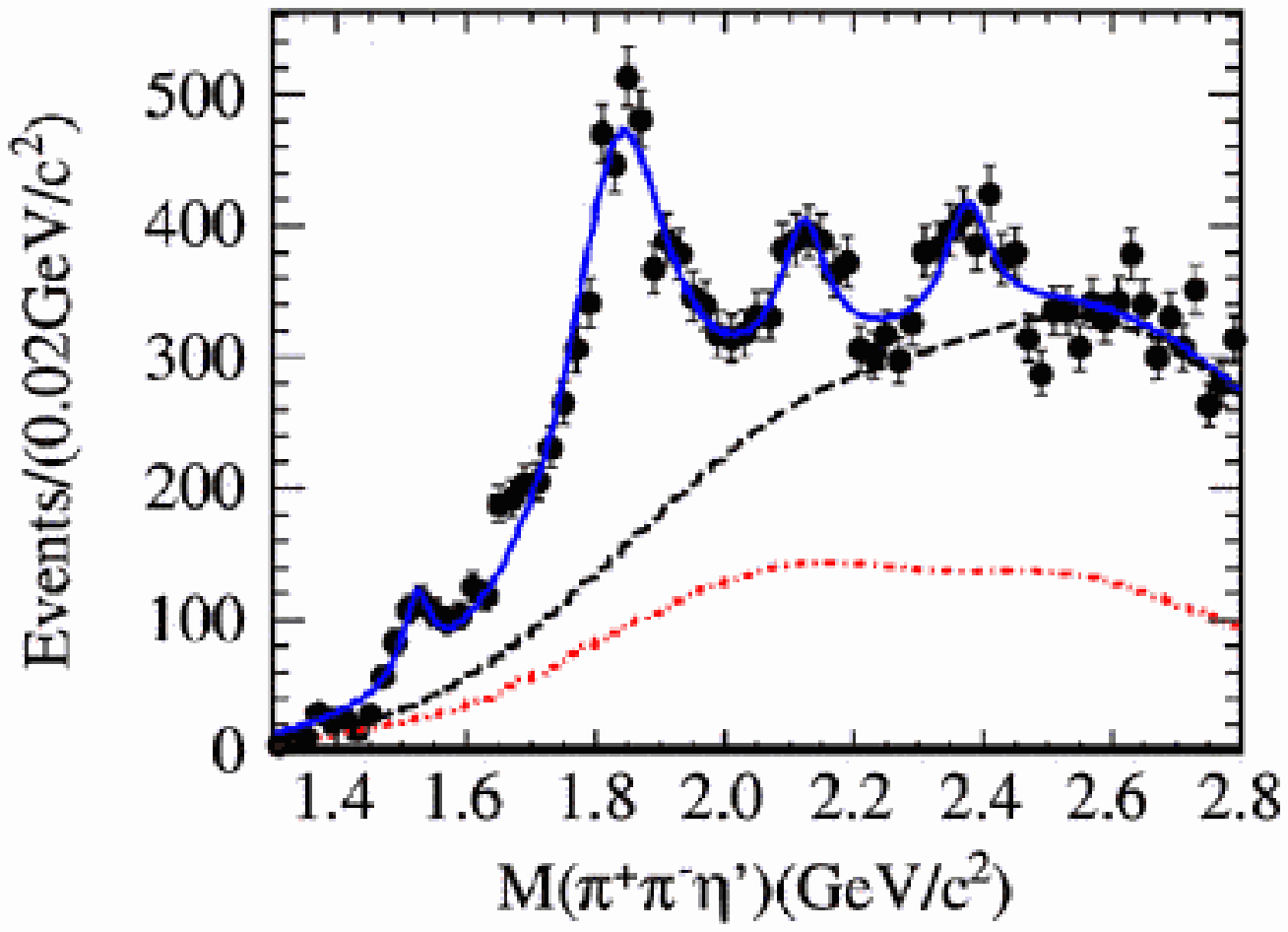}}
\subfigure[]{\label{fig:1c}\includegraphics[width=0.3\textwidth,height=0.27\textwidth]{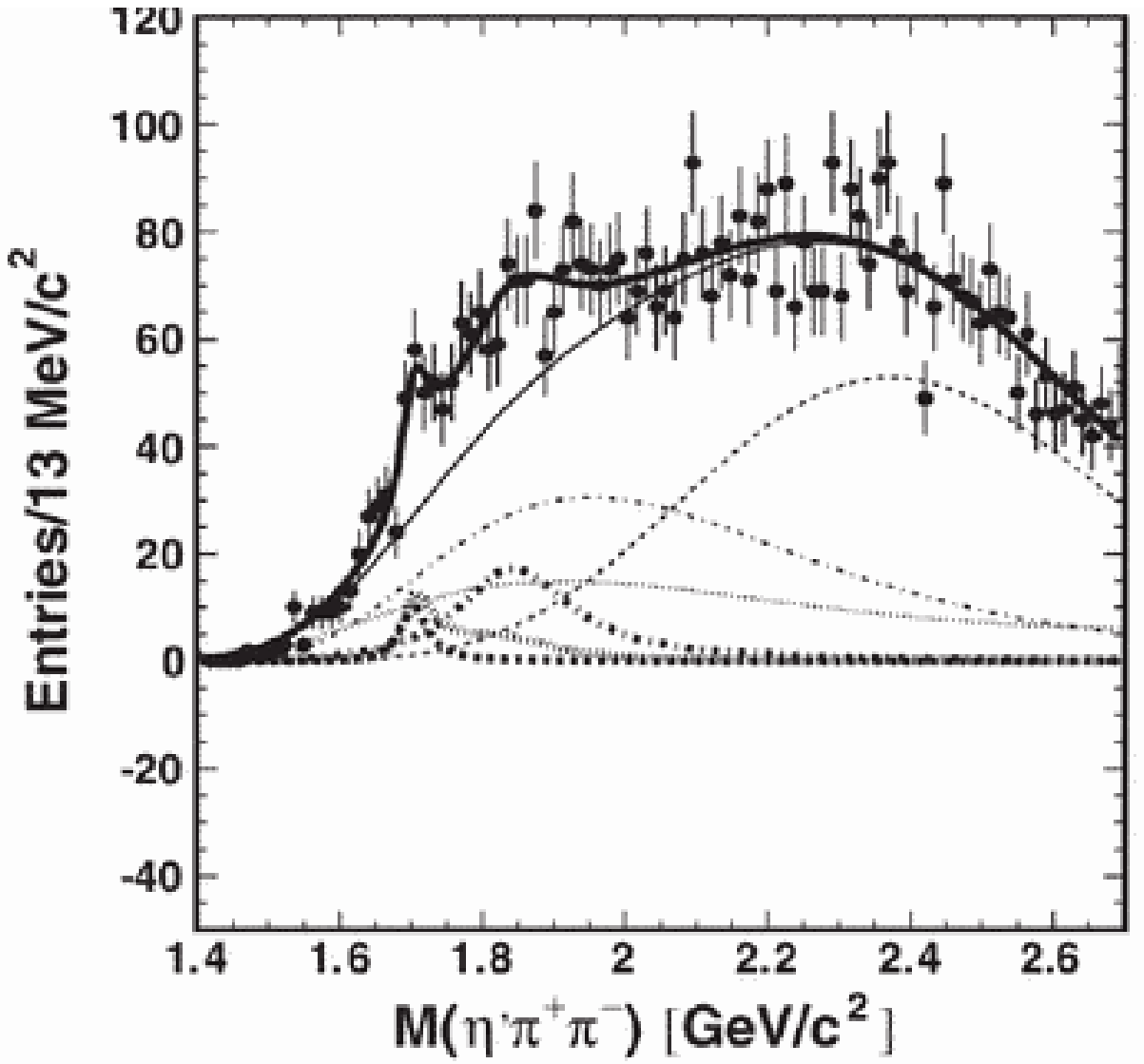}}
\subfigure[]{\label{fig:1d}\includegraphics[width=0.3\textwidth,height=0.27\textwidth]{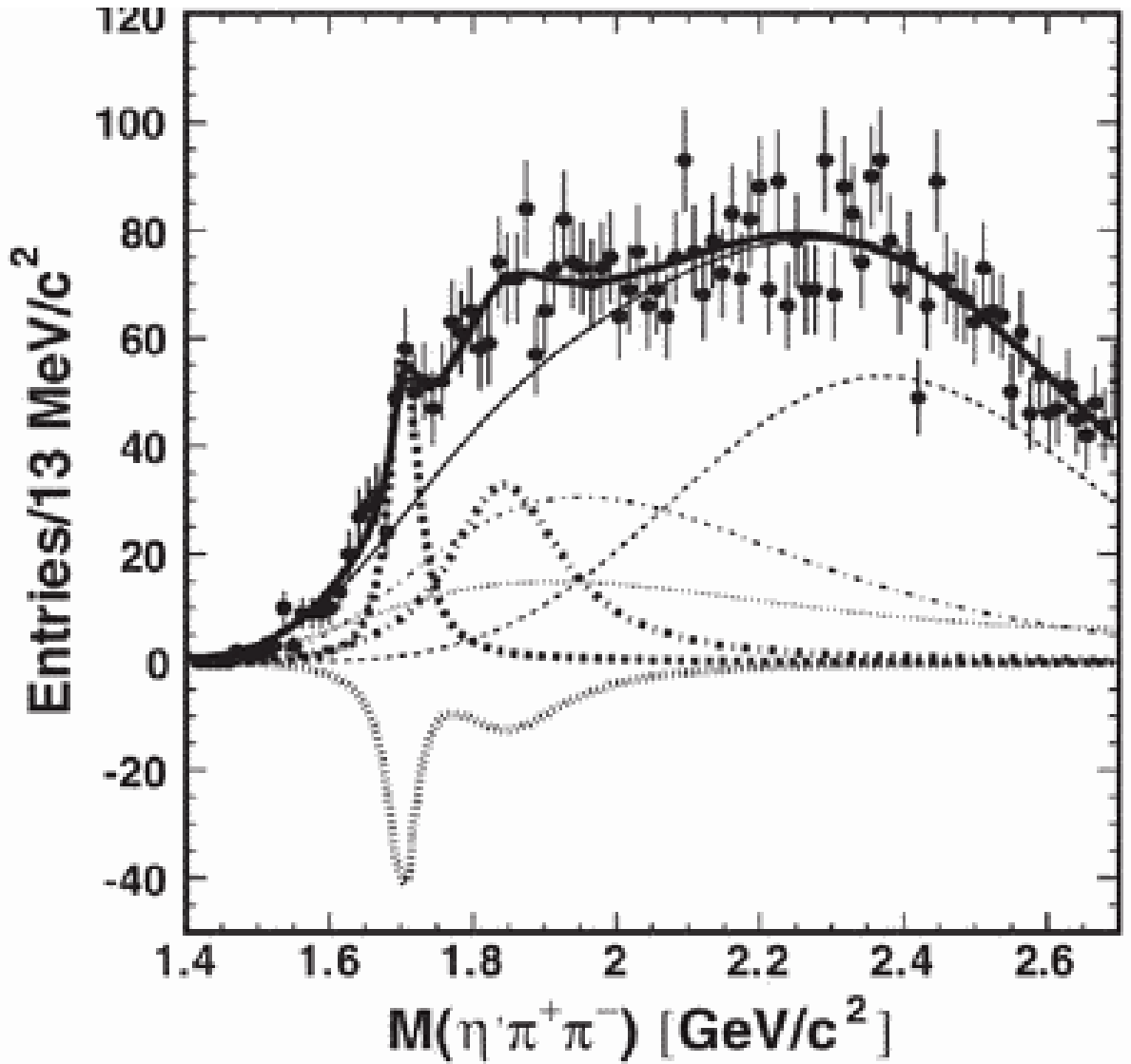}}
\subfigure[]{\label{fig:1e}\includegraphics[width=0.3\textwidth,height=0.27\textwidth]{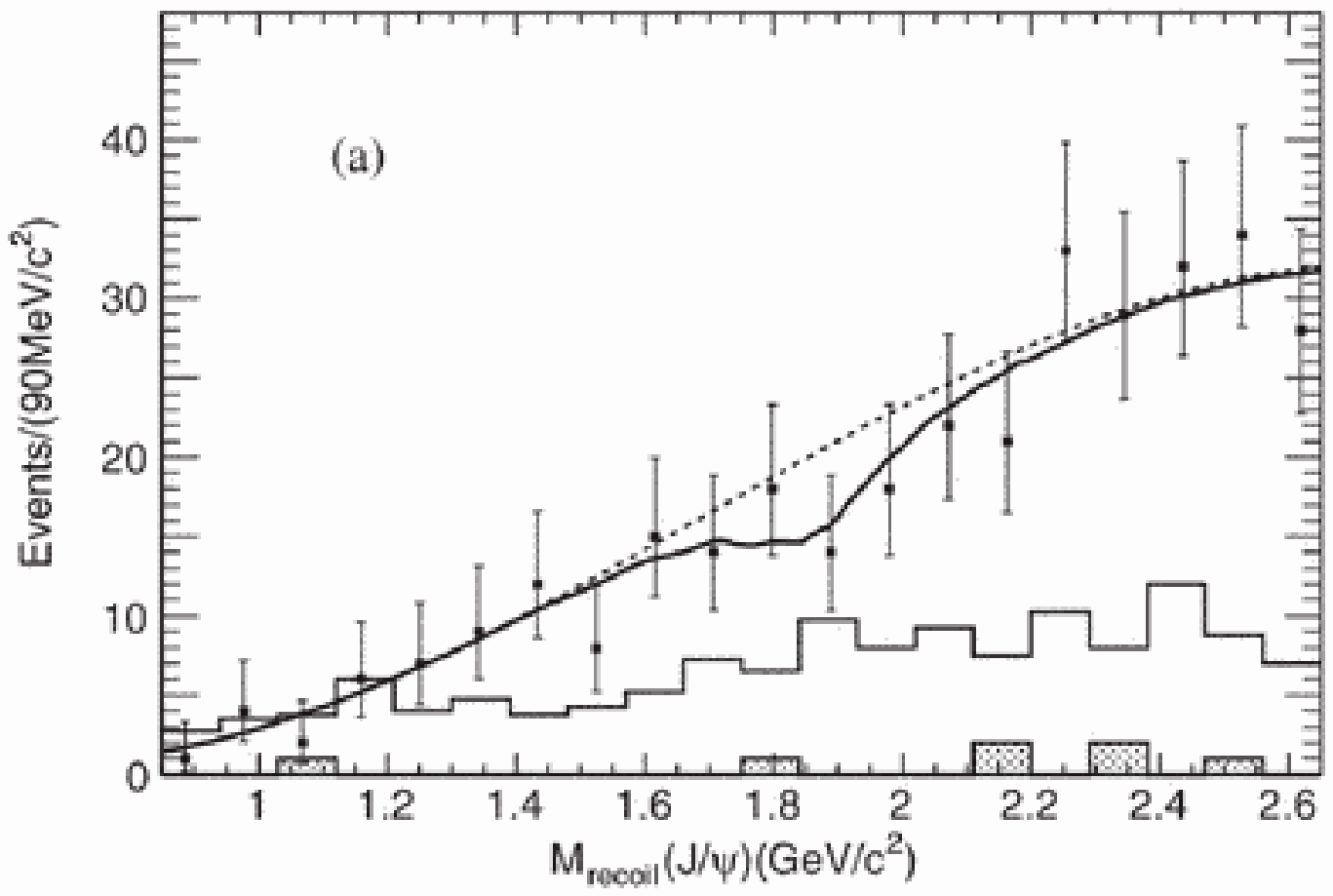}}
\subfigure[]{\label{fig:1f}\includegraphics[width=0.3\textwidth,height=0.27\textwidth]{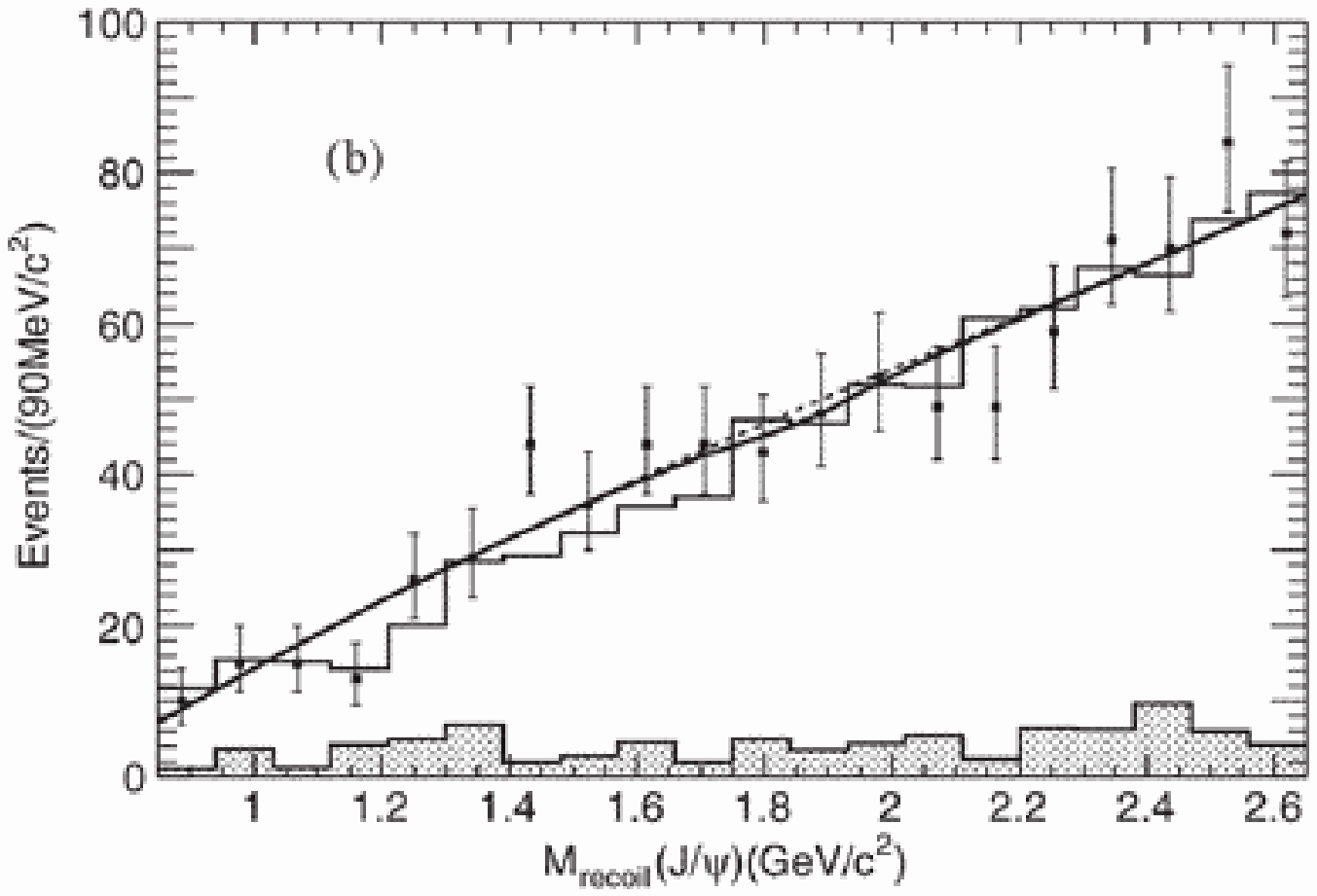}}
\subfigure[]{\label{fig:1g}\includegraphics[width=0.3\textwidth,height=0.27\textwidth]{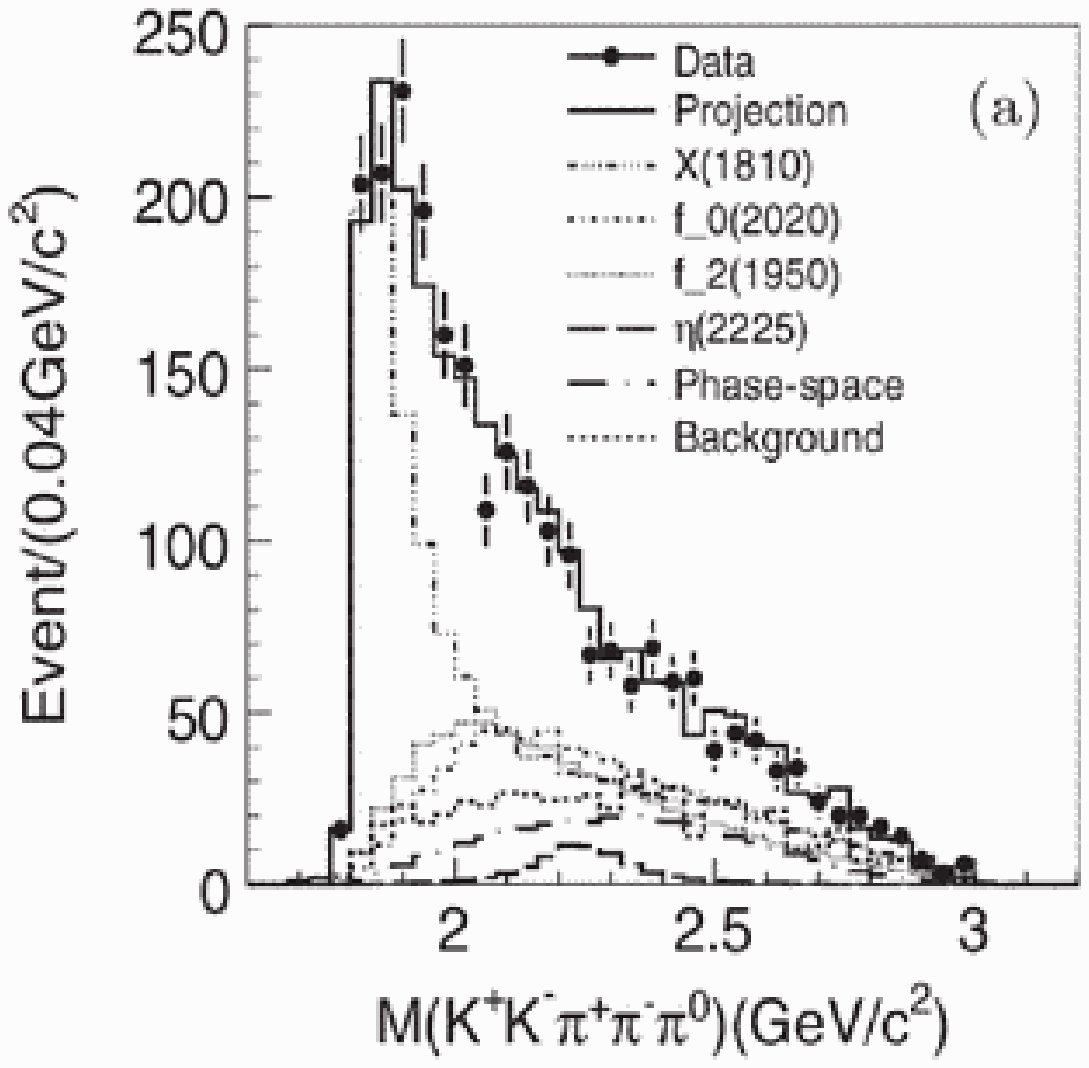}}
\subfigure[]{\label{fig:1h}\includegraphics[width=0.3\textwidth,height=0.27\textwidth]{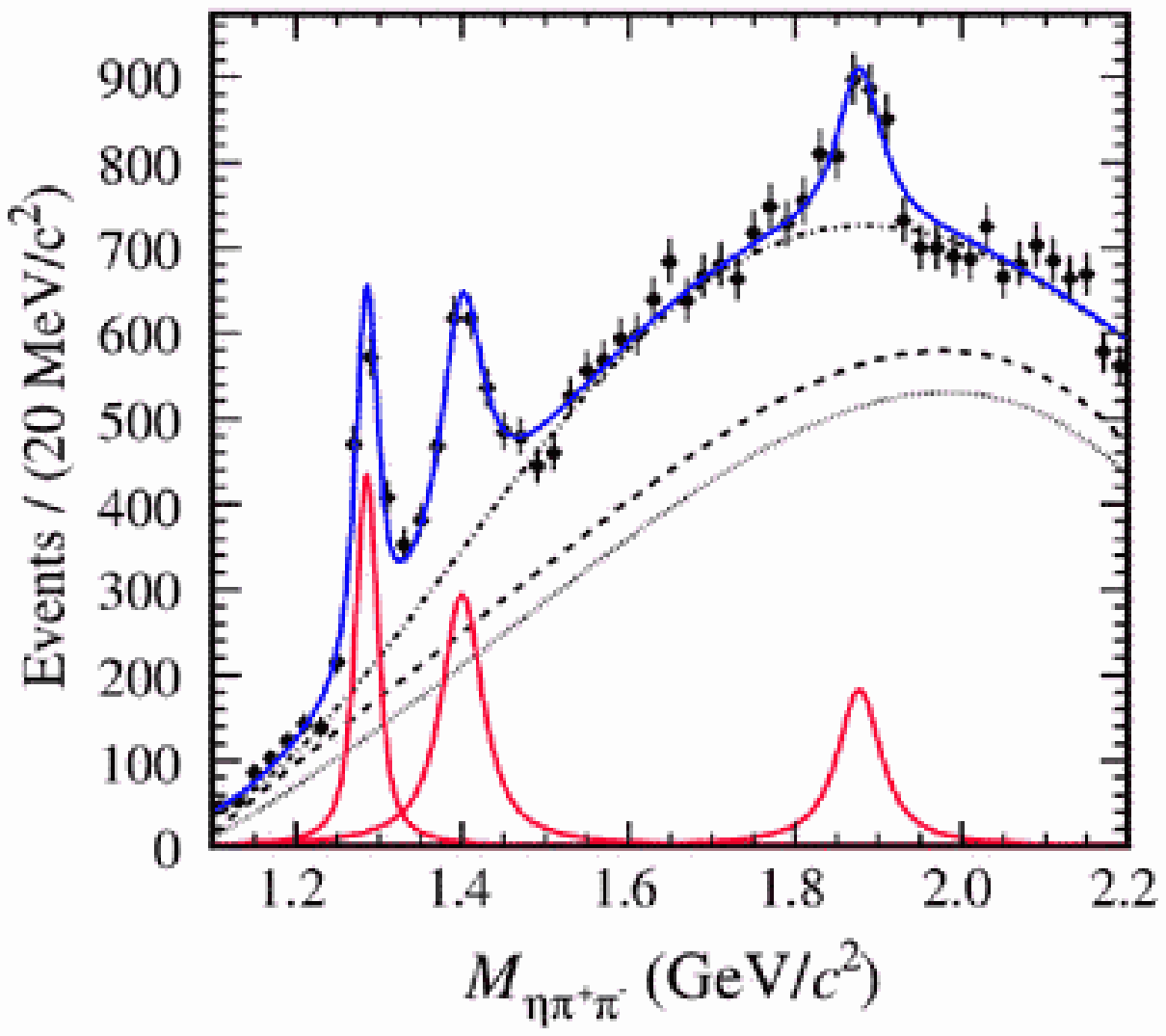}}
\subfigure[]{\label{fig:1i}\includegraphics[width=0.3\textwidth,height=0.27\textwidth]{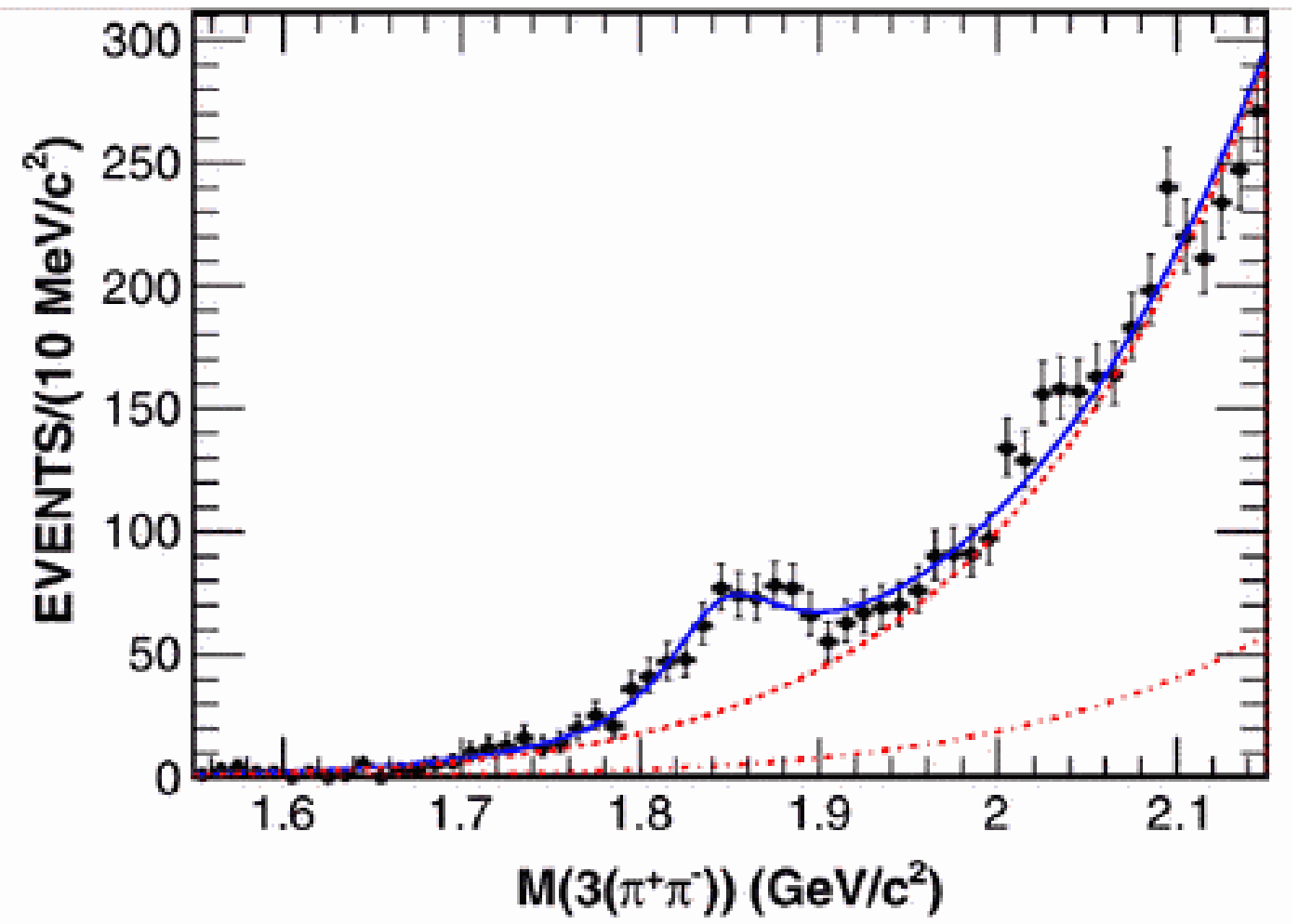}}
\caption{Structures near the \ppbar threshold via different processes. (a):
  $J/\psi \to \gamma p \bar{p}$; (b): $J/\psi \to \gamma
\pi^+ \pi^- \eta'$; (c): $\gamma \gamma \to \pi^+ \pi^- \eta'$ (constructive); (d):
  $\gamma \gamma \to \pi^+ \pi^- \eta'$ (destructive); (e): $e^+ e^- \to J/\psi + X(1835)$
  ($e^+e^-$ mode); (f): $e^+ e^- \to J/\psi + X(1835)$ ($\mu\mu$ mode); (g): $J/\psi \to
  \gamma \omega \phi$; (h): $J/\psi \to \omega a^\pm_0(980) \pi^\mp$; (i): $J/\psi \to \gamma 3(\pi^+\pi^-)$.}
\label{fig:thresholds}
\end{figure}

   An anomalous enhancement near the \ppbar~ mass-threshold was first observed by the BESII
experiment in the $J/\psi$ radiative decay process $J/\psi\to \gamma p
\bar{p}$~\cite{PhysRevLett.91.022001} and was confirmed by the
BESIII~\cite{ChinPhys.C34.421} and CLEO-c~\cite{PhysRev.D82.092002} experiments. Recently,
a partial wave analysis (PWA) in the radiative decay $J/\psi \to \gamma p \bar{p}$ is
applied~\cite{PhysRevLett.108.112003}, the $J^{PC}$ quantum numbers of the \ppbar~
mass-threshold enhancement is determined to be $0^{-+}$, with peak mass $M=
1832^{+19}_{-5}(stat)^{+18}_{-17}(syst)\pm 19 (model) \mathrm{MeV/c^2}$ below the
threshold and total width $\Gamma = 13 \pm 39(stat)^{+10}_{-13}(syst)\pm 4 (model)
\mathrm{MeV/c^2}$ at the $90\%$ C.L. The product of branching fractions is $Br[J/\psi\to
\gamma X(p\bar{p})]Br[X(p\bar{p}) \to p\bar{p}] =
[9.0^{+0.4}_{-1.1}(stat)^{+1.5}_{-5.0}(sys)\pm2.3(model)]\times
10^{-5}$. Fig.~\ref{fig:1a} shows a comparison between data and PWA fit projection on the
\ppbar~invariant mass; here, the black dots with error bars are data, the solid histograms
show the PWA total projection, and the dashed, dotted, dash-dotted, and dash-dot-dotted
lines show the contributions of the $X(p\bar{p})$, $0^{++}$ phase space, $f_0(2100)$ and
$f_2(1910)$, respectively.

However, such near \ppbar~ mass-threshold structure has not been observed in \ppbar~ cross
section measurements, in B-meson decays~\cite{IntJModPhys.A20.5145,PhysRevLett.92.131801},
in radiative $\psi(3686)$ or $\Upsilon$ decays to \ppbar~\cite{PhysRevLett.99.011802,
PhysRev.D83.032001}, or in $J/\psi\to \omega p \bar{p}$ decays~\cite{arxiv:0710.5369,
PhysRev.D87.112004}. These non-observations exclude the attribution of the mass-threshold
enhancement to a pure final state interaction
(FSR)~\cite{PhysRev.D71.054010,PhysLett.B692.136,PhysRev.D69.034004,PhysRev.D80.034032,
PhysLett.B633.283, PhysRev.D73.014023}.

Inspired by the anomalous \ppbar~invariant mass threshold enhancement, $J/\psi \to \gamma
\pi^+ \pi^- \eta'$ is searched and a $\pi^+ \pi^- \eta'$ resonance, the $X(1835)$, was
observed by the BESII experiment~\cite{PhysRevLett.95.262001}. Recently, with a larger
\jpsi~sample, $(225.2\pm2.8)\times 10^6$ events registered in the BESIII detector, $J/\psi
\to \gamma \pi^+ \pi^- \eta'$ is studied using two $\eta'$ decay modes: $\eta' \to \pi^+
\pi^- \eta$ and $\eta' \to \gamma \rho^0$~\cite{PhysRevLett.106.072002}. The $X(1835)$ is
confirmed with a statistical significance larger than $20$ standard deviations, as well as
two new structures, the $X(2120)$ and $X(2370)$, are observed in the $\pi^+ \pi^- \eta'$
invariant-mass spectrum with statistical significances larger than $7.2\sigma$ and
$6.4\sigma$, respectively. Fig.~\ref{fig:1b} shows the mass spectrum fitted with four
resonances; here, the dash-dotted line is contributions of non-$\eta'$ events and the
$\pi^0\pi^+\pi^-\eta'$ background for two $\eta'$ decay modes, and the dashed line is
contributions of the total background and non-resonant $\pi^+\pi^-\eta'$ process. The
masses and widths of $X(1835)$ are measured to be $M=1836.5 \pm 3.0
(stat)^{+5.6}_{-2.1}(syst)\ \mathrm{MeV/c^2}$, $\Gamma = 190 \pm
9(stat)^{+38}_{-36}(syst)\ \mathrm{MeV/c^2}$. The product branching ratios is $Br[J/\psi\to
\gamma X(1835)]\cdot Br[X(1835)\to \pi^+ \pi^- \eta'] = [2.87 \pm 0.09
(stat)^{+0.49}_{-0.52}(syst)] \times 10^{-4}$, and the angular distribution of the
radiative photon is consistent with a pseudo-scalar assignment. Another attempt to search
$X(1835)$ via $\gamma \gamma \to \pi^+ \pi^- \eta'$ is implemented by
Belle~\cite{PhysReV.D86.052002}, in which the interference between $X(1835)$ and
$\eta(1760)$ has been considered, but no strong evidence was found. Figs.~\ref{fig:1c} and
~\ref{fig:1d} show the constructive and destructive interference respectively. Inspired by
a C-even glueball can be studied in the process $e^+ e^- \to \gamma^* \to H
G$~\cite{PhysRevLett.91.112001}, where $H$ denotes a $c\bar{c}$ quark pair or a charmonium
state and $G$ is a glueball, $X(1835)$ is also searched via $e^+ e^- \to J/\psi + x(1835)$
at $\sqrt{s}\approx 10.6 \mathrm{GeV}$~\cite{PhysRev.D89.032003}. Figs.~\ref{fig:1e}
and~\ref{fig:1f} show no significant evidence is found to support the hypothesis of the
$X(1835)$ as a glueball produced in association with a $J/\psi$.

An anomalous near-threshold enhancement in the $\omega \phi$ invariant-mass spectrum in
the process $J/\psi \to \gamma \omega \phi$ was reported by the BESII
experiment~\cite{PhysRevLett.96.162002}, that was assumed to an existence of a resonance,
\ie the $X(1810)$.  With the new $J/\psi$ event sample accumulated with the BESIII
detector, the process is re-studied recently~\cite{PhysRev.D87.032008}, and a partial
wave analysis with a tensor co-variant amplitude is performed. The projection on the
invariant mass of $\omega \phi$ is shown in Fig.~\ref{fig:1g}. The spin-parity of the
$X(1810)$ is determined to be $0^{++}$. Its mass and width are measured to be $M=1795 \pm
7 (stat)^{+13}_{-5}(syst)\pm 19(mode)\mathrm{MeV/c^2}$ and $\Gamma = 95 \pm
10(stat)^{+21}_{-34}(syst) \pm 75(mode)\mathrm{MeV/c^2}$, respectively. And the product
branching fraction is determined to be $Br[J/\psi\to\gamma X(1810)]\times Br[X(1810) \to
  \omega \phi] = [2.00 \pm 0.08(stat) ^{+0.45}_{-1.00}(syst)\pm 1.30 (mod)]\times
10^{-4}$. 

Using the same data sample as mentioned above, BESIII collaboration reported the
observation of a new process $J/\psi \to \omega X(1870)$~\cite{PhysRevLett.107.182001}
with a statistical significance of $7.2\sigma$, in which $X(1870)$ decays to $a^\pm_0(980)
\pi^\mp$. Fitting to $\eta \pi^+\pi^-$ mass spectrum yields the mass, width and product
branching fraction are $M=1877.3 \pm
6.3(stat)^{+3.4}_{-7.4}(syst) \mathrm{MeV/c^2}$, $\Gamma = 57 \pm 12 (stat)^{+19}_{-4}
(syst) \mathrm{MeV/c^2}$, and $Br(J/\psi \to \omega X)\times Br(X\to a^\pm_0(980) \pi^\mp)
\times Br(a^\pm_0(980) \to \eta \pi^\pm) = [1.50 \pm
  0.26(stat)^{+0.72}_{-0.36}(syst)]\times 10^{-4}$. Fig.~\ref{fig:1h} shows the results of
the fit to the $M(\eta \pi^+ \pi^-)$ mass distribution for the events with either the
$\eta \pi^+$ or $\eta \pi^-$ in the $a_0(980)$ mass window. The dotted curve shows the
contribution of non-$\omega$ and/or non-$a_0(980)$ background, the dashed line also
includes the contribution from $J/\psi \to b_1(1235)a_0(980)$, and the dot-dashed curve
indicates the total background with the non-resonant $J/\psi \to \omega
a^\pm_0(980) \pi^\mp$ included. $\chi^2/d.o.f.$ is $1.27$ for this fit.

BESIII also analyzed the decay $J/\psi \to \gamma 3(\pi^+\pi^-)$ with the same $J/\psi$
sample~\cite{PhysRev.D88.091502}. A structure at $1.84\mathrm{GeV/c^2}$ is observed in the
$3(\pi^+\pi^-)$ invariant mass spectrum with a statistical significance of
$7.6\sigma$. The mass and width are measured to be $M=1842.2 \pm 4.2^{+7.1}_{-2.6}
\mathrm{MeV/c^2}$ and $\Gamma = 83 \pm 14 \pm 11 \mathrm{MeV}$. The product branching
fraction is determined to be $Br[J/\psi \to \gamma X(1840)] \times Br[X(1840)\to
  3(\pi^+\pi^-)] = (2.44 \pm 0.36^{+0.60}_{-0.74})\times 10^{-5}$. Fig.~\ref{fig:1i} shows
the fit of the mass spectrum of $3(\pi^+\pi^-)$. The dots with error bars are data; the
solid line is the fit result. The dashed line represents all the backgrounds, including
the background events from $J/\psi \to \pi^0 3(\pi^+ \pi^-)$ (represented by the
dashed-dotted line, fixed in the fit) and a third-order polynomial representing other
backgrounds. 

\begin{figure}[ht!]
\centering
\includegraphics[width=0.8\textwidth]{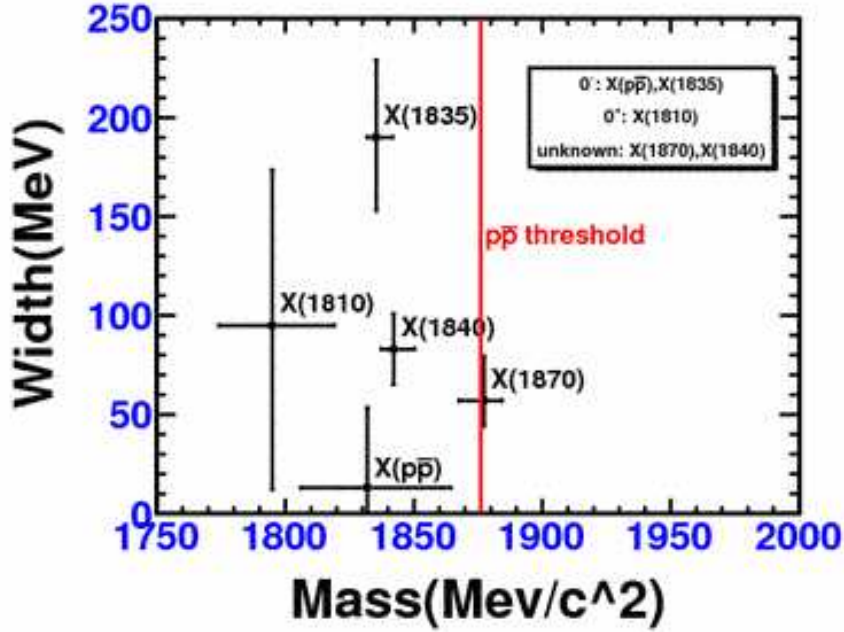}
\caption{Comparison of masses, widths and quantum numbers of the structures near the $p
\bar{p}$ threshold.}
\label{fig:threshold:compare}
\end{figure}

Fig.~\ref{fig:threshold:compare} shows the comparison of the masses and widths of all the
structures near threshold of \ppbar mentioned before. Four of the five are from $J/\psi$
radiative decays. Although it's well known that $J/\psi$'s radiative decays favor
glue-balls, we are reasonably sure that $X(1835)$ would not be a glueball since it has not
been observed in a production associated with a $J/\psi$. We also know that $X(p\bar{p})$
is not found in $\psi'$ radiative decay and $J/\psi \to \omega p \bar{p}$, then it does
not likely from a pure FSI.  A number of theoretical speculations have been proposed to
interpret the nature of the structure near the \ppbar~threshold
structure~\cite{PhysRev.D71.054010, PhysLett.B692.136, PhysRev.D69.034004,
PhysRev.D80.034032, PhysLett.B633.283, PhysRev.D73.014023, PhysLett.B567.273,
PhysReV.D72.034027, PhysRev.C72.011001}. Among them, the most intriguing suggestion is
that it is due to a \ppbar~bound state, also called baryonium~\cite{PhysLett.B567.273,
PhysReV.D72.034027, PhysRev.C72.011001}, a predicted state with a long history and been
searched by many experiments~\cite{PhysRept.368.119}. However, all the information we
obtained till now is too limited to draw a final conclusion. We even don't know whether
all, or part of, the enhancements near the \ppbar~threshold are related to a same source
still. Further studies are surely needed; among these, spin-parity determination,
precise measurements of masses, widths, and branching ratios are especially important.

\section{XYZ particles}
\begin{figure}[ht!]
\centering
\includegraphics[width=0.8\textwidth]{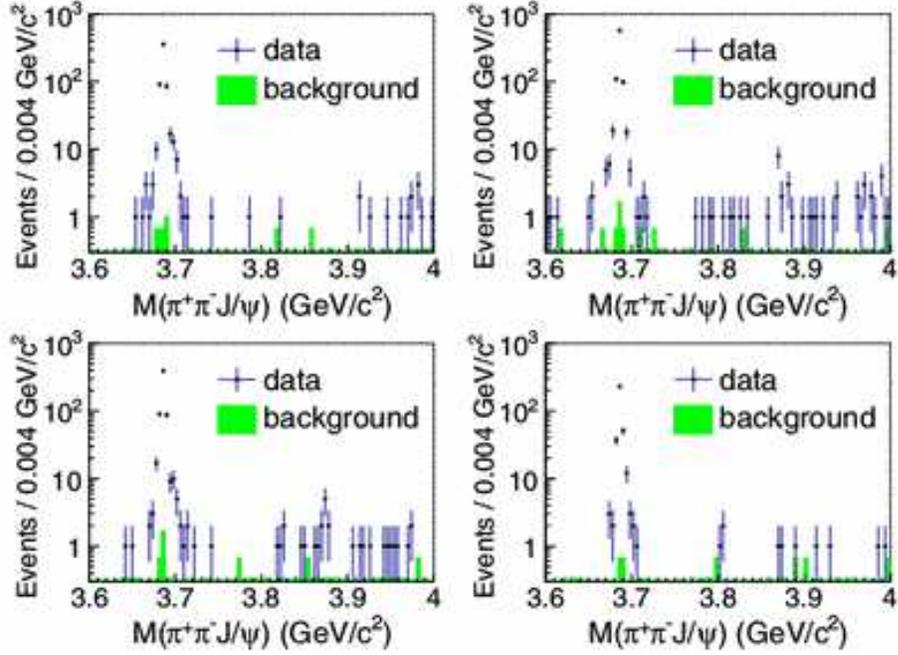}
\caption{The $\pi^+ \pi^- J/\psi$ invariant mass distributions at $\sqrt{s} = 4.009$ (top
  left), $4.229$ (top right), $4.260$ (bottom left), and $4.360$ $\mathrm{GeV}$ (bottom
  right). Dots with error bars are data, the green shaded histograms are normalized
  $J/\psi$ sideband events.}
\label{fig:x3872}
\end{figure}

$X(3872)$ was first observed in the process $B\to K(\pi^+ \pi^- J/\psi)$ by
Belle~\cite{PhysRevLett.91.262001}, its mass is close to the $D^0 D^{*0}$ threshold and
width is very narrow. CDF~\cite{PhysRevLett.98.132002} and
LHCb~\cite{PhysRevLett.110.222001} determined its $J^{PC}$ is $1^{++}$. The partial width
measurement shows that it takes a $50\%$ chance to decay via open-charm channel and at the
$O(\%)$ via charmonium. There are many proposed interpretation for its nature since
its discovery, such as $\chi_{c1}(2P)$, $D^0 D^{*0}$ module, hybrid of
meson and glue-ball, and tetra-quark states, \etal. However none of them is totally
satisfactory and the nature of $X(3872)$ is still a mystery. $X(3872)$ has only been found
in the pp collision and B decays, until recently BESIII reported a new production mode of
$e^+ e^- \to \gamma (\pi^+ \pi^- J/\psi)$~\cite{PhysRevLett.112.092001}, with data samples
collected with the BESIII detector at center-of-mass energies from $4.009$ to
$4.420~\mathrm{GeV}$. Figs.~\ref{fig:x3872} show the $\pi^+ \pi^- J/\psi$ invariant mass
distributions at four energies. The measurements results are consistent with expectations
for the radiative transition process $Y(4260) \to \gamma X(3872)$.

\begin{figure}[ht!]
\centering
\subfigure[]{\label{fig:ya}\includegraphics[width=0.3\textwidth,height=0.27\textwidth]{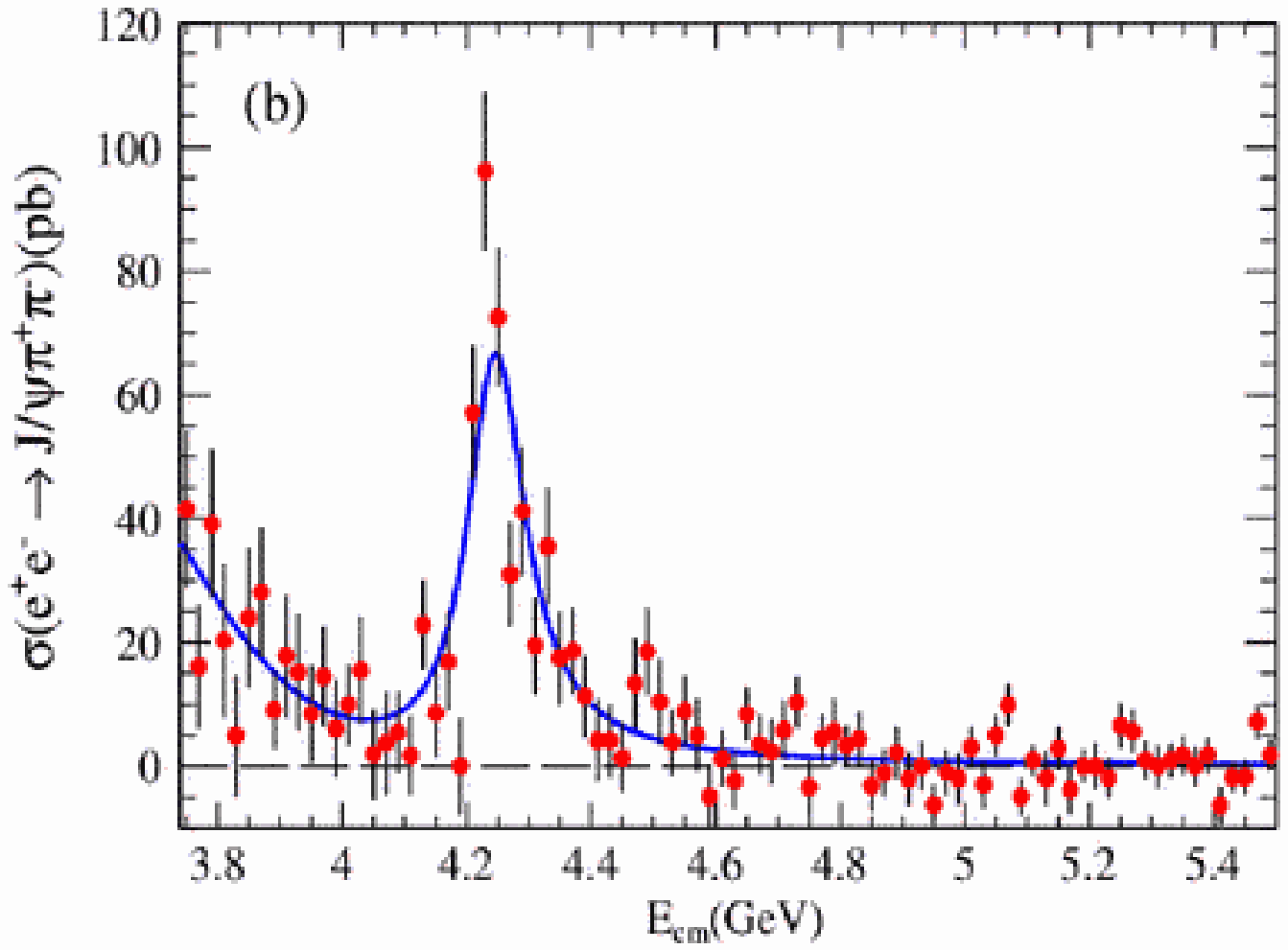}}
\subfigure[]{\label{fig:yb}\includegraphics[width=0.3\textwidth,height=0.27\textwidth]{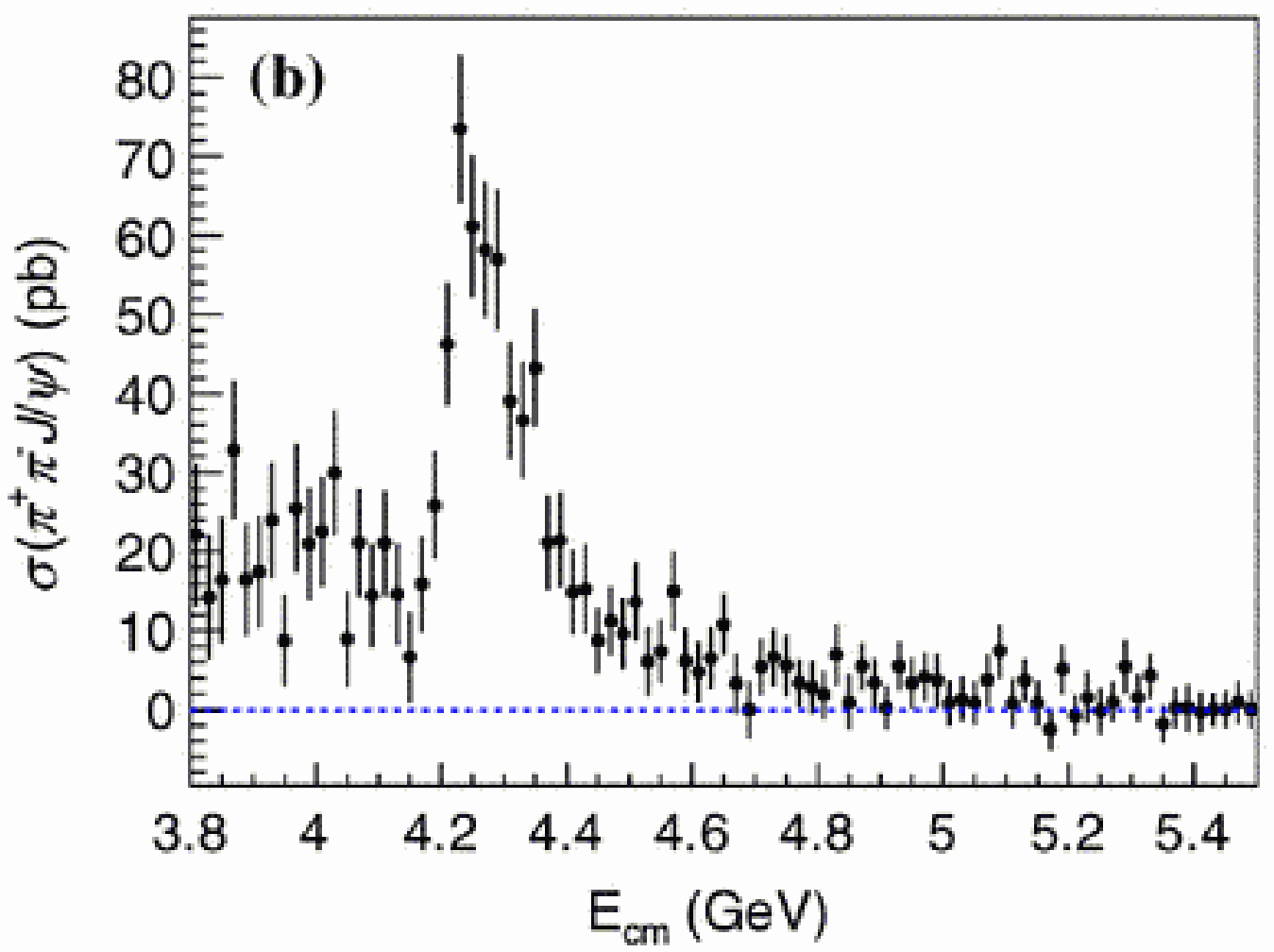}}
\subfigure[]{\label{fig:yc}\includegraphics[width=0.3\textwidth,height=0.27\textwidth]{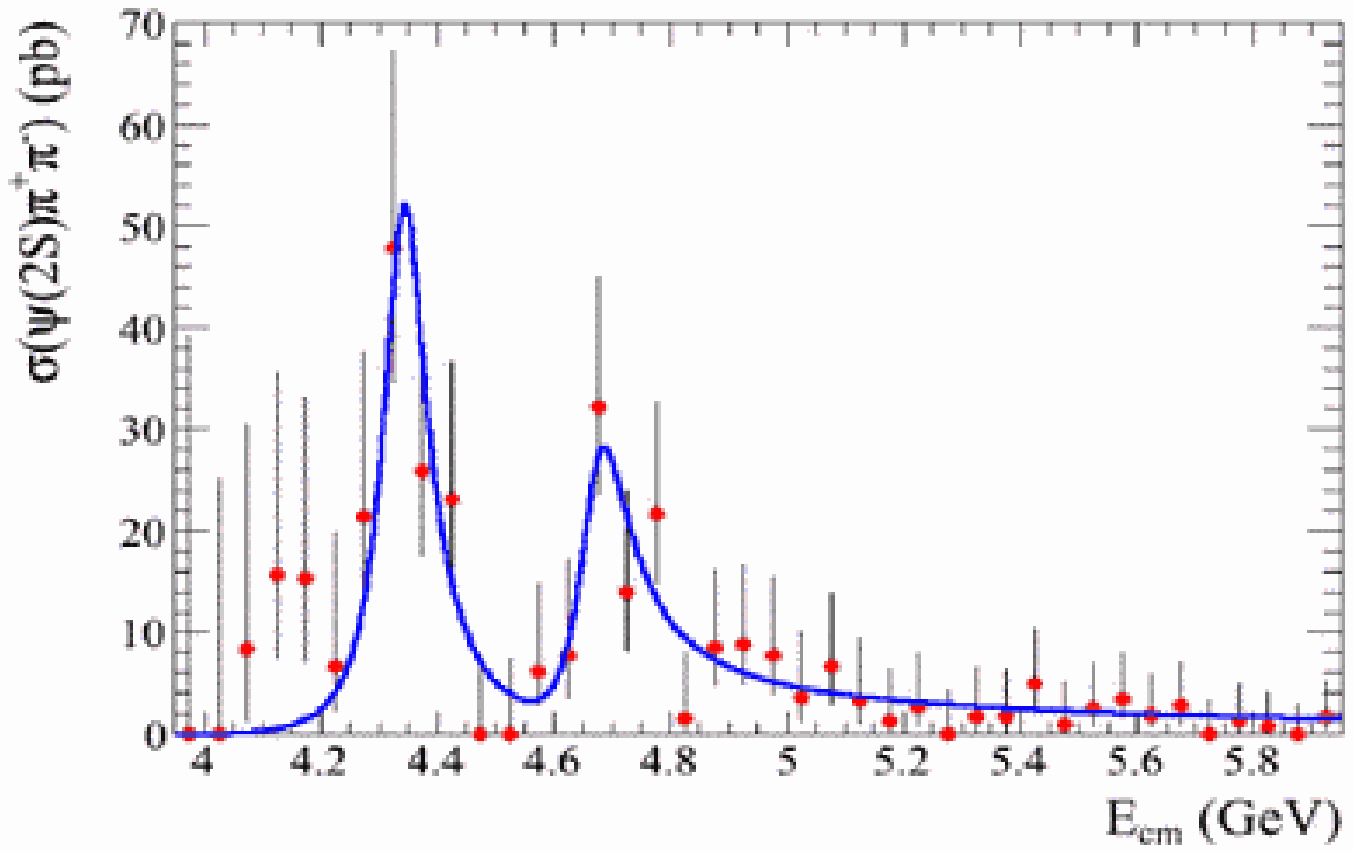}}
\caption{Cross section of $\pi^+ \pi^- J/\psi$ and $\pi^+ \pi^- \psi(2S)$ final
  states. (a) $\pi^+ \pi^- J/\psi$ mode from Babar; (b) $\pi^+ \pi^- J/\psi$ mode from
  Belle; (c) $\pi^+ \pi^- \psi(2S)$ mode from Babar.}
\label{fig:ystates}
\end{figure}

Many, maybe too many, Y states have been reported in last decade, such as $Y(4008)$,
$Y(4260)$, $Y(4360)$, $Y(4630)$, and $Y(4660)$, \etal. Most of them are found in ISR
processes, so they are vector states, where the total number of them is surely beyond the
theoretical prediction of charmonium states at this energy region by the potential
model. One interesting thing is these Y states are only discovered in $\pi^+ \pi^-
J/\psi$~\cite{PhysRevLett.95.142001,PhysRevLett.99.182004}, $\pi^+\pi^-
\psi(2S)$~\cite{PhysRevLett.99.142002}, and $\Lambda^+_c
\Lambda^-_c$~\cite{PhysRevLett.101.172001} modes, but there is no sign of $Y\to
D^{(*)}D^{(*)}$~\cite{PhysRev.D77.011103, PhysRevLett.98.092001, PhysRevLett.100.062001}
modes. Recently, Babar and Belle have updated their studies of Y states with larger
statistics. Figs.~\ref{fig:ya} and~\ref{fig:yb} show the cross sections of $\pi^+ \pi^-
J/\psi$ final states from Babar~\cite{PhysRev.D86.051102} with $454 fb^{-1}$ and
Belle~\cite{PhysRevLett.110.252002} with $967 fb^{-1}$ electron-positron colliding events
via ISR process. In Belle's report, still two resonances $Y(4008)$ and $Y(4260)$ are
observed, just agrees with Belle's previous results. In Babar's result, significant
$Y(4260)$ is found, but $Y(4008)$ is not confirmed. But from Fig.~\ref{fig:ya}, there are
some events accumulated at around the $4.008$ GeV, so it may due to different fit methods
used by Babar and Belle. For the $e^+ e^- \to \gamma_{ISR} \pi^+ \pi^- \psi(2S)$, Babar
updated it with $520 pb^{-1}$~\cite{arxiv:1211.6271} and Belle with $980
fb^{-1}$. Figs.~\ref{fig:yc} shows the cross section of $\pi^+ \pi^- \psi(2S)$ final
states from Babar, and it is fitted with $Y(4360)$ and $Y(4660)$ two resonances, and two
solutions are found. As a preliminary result, Belle fitted its data with $Y(4260)$,
$Y(4360)$ and $Y(4660)$, it turns out the significance of $Y(4260)$ is only $2.1\sigma$,
but its effect on the others is large.

\begin{figure}[ht!]
\centering
\subfigure[]{\label{fig:ye}\includegraphics[width=0.3\textwidth,height=0.27\textwidth]{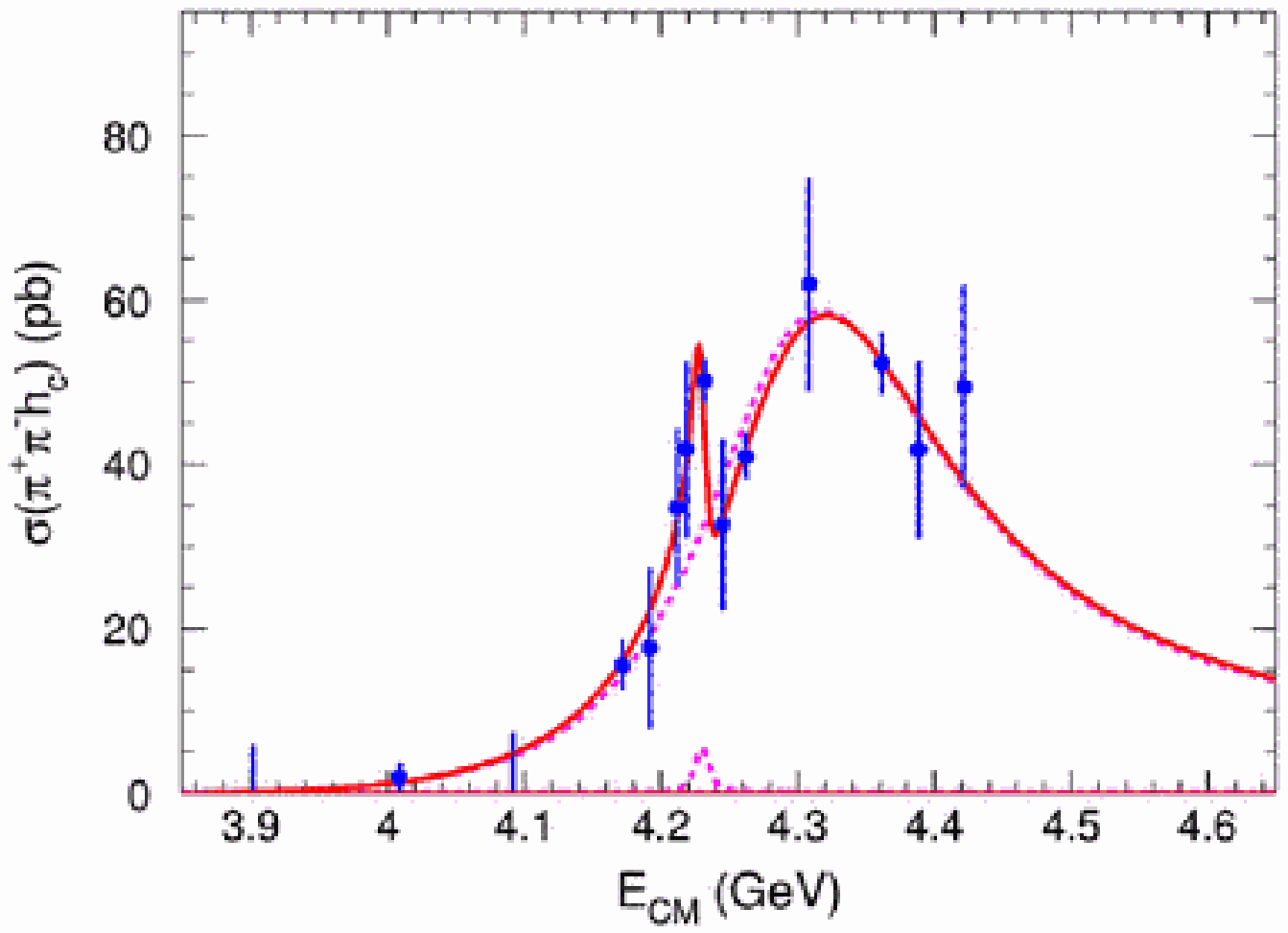}}
\subfigure[]{\label{fig:yf}\includegraphics[width=0.3\textwidth,height=0.27\textwidth]{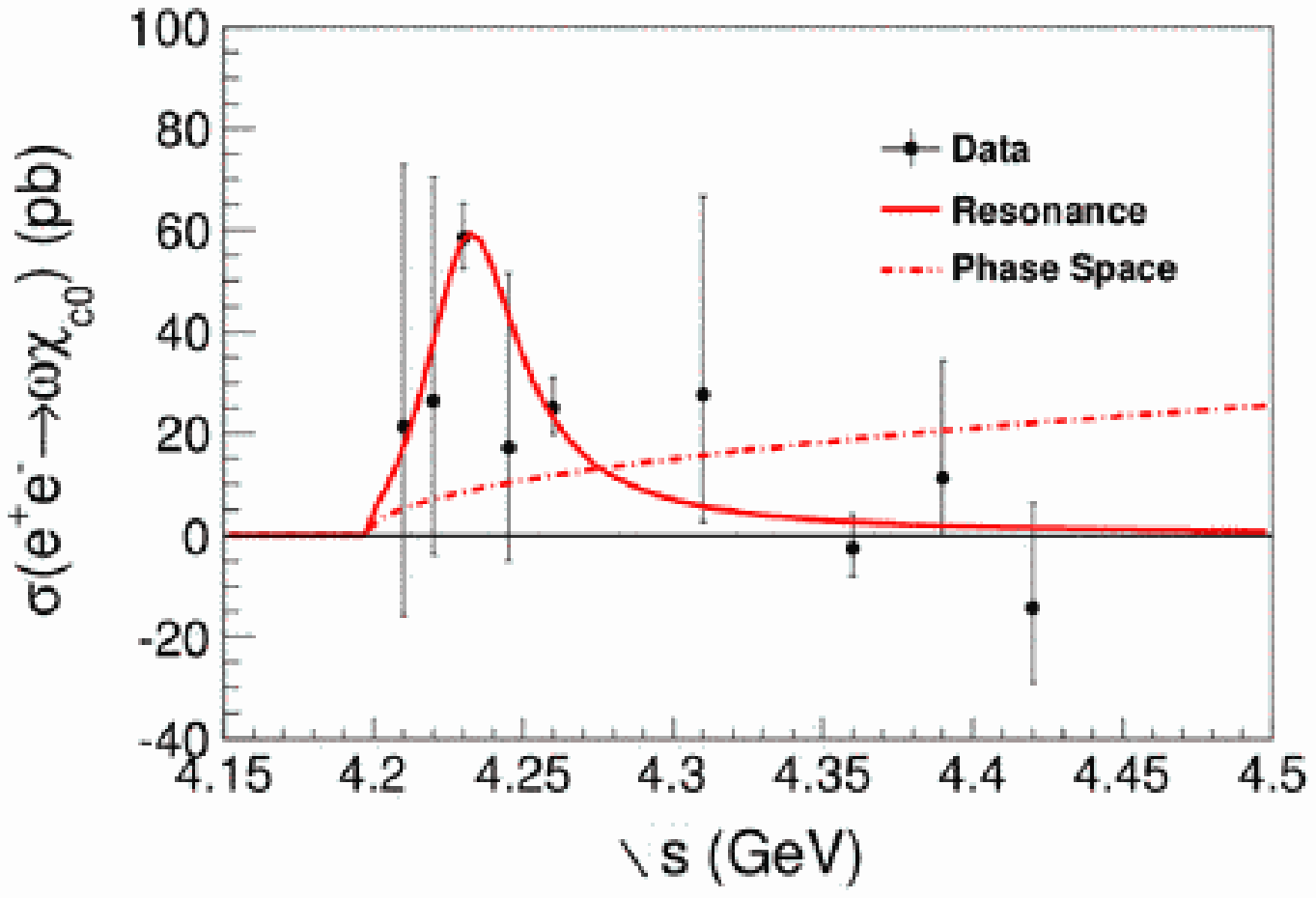}}
\caption{(a) Fit based on the combined BESIII and CLEO-c data; (b) Cross section of $\omega \chi_{c0}$.}
\label{fig:ystates2}
\end{figure}

There are also other attempts to study the Y states via $\pi^+ \pi^- h_c$
~\cite{PhysRevLett.107.041803, PhysRevLett.111.242001} and $\omega \chi_{cJ}$ final
states. Fig.~\ref{fig:ye} shows an individual work based on the combined BESIII and CLEO-c
data~\cite{arxiv:1310.0280}, where a narrow $Y(4220)$ and a wide $Y(4290)$ states are used
to describe the structures. Fig.~\ref{fig:yf} shows a cross section distribution of
$\omega \chi_{c0}$ as a preliminary result of BESIII, and no signal of $\omega \chi_{c1}$
and $\omega \chi_{c2}$ are found disfavors $Y(4260)$ is a $\omega \chi_{c1}$ module.

\begin{figure}[ht!]
\centering
\subfigure[]{\label{fig:zca}\includegraphics[width=0.3\textwidth,height=0.27\textwidth]{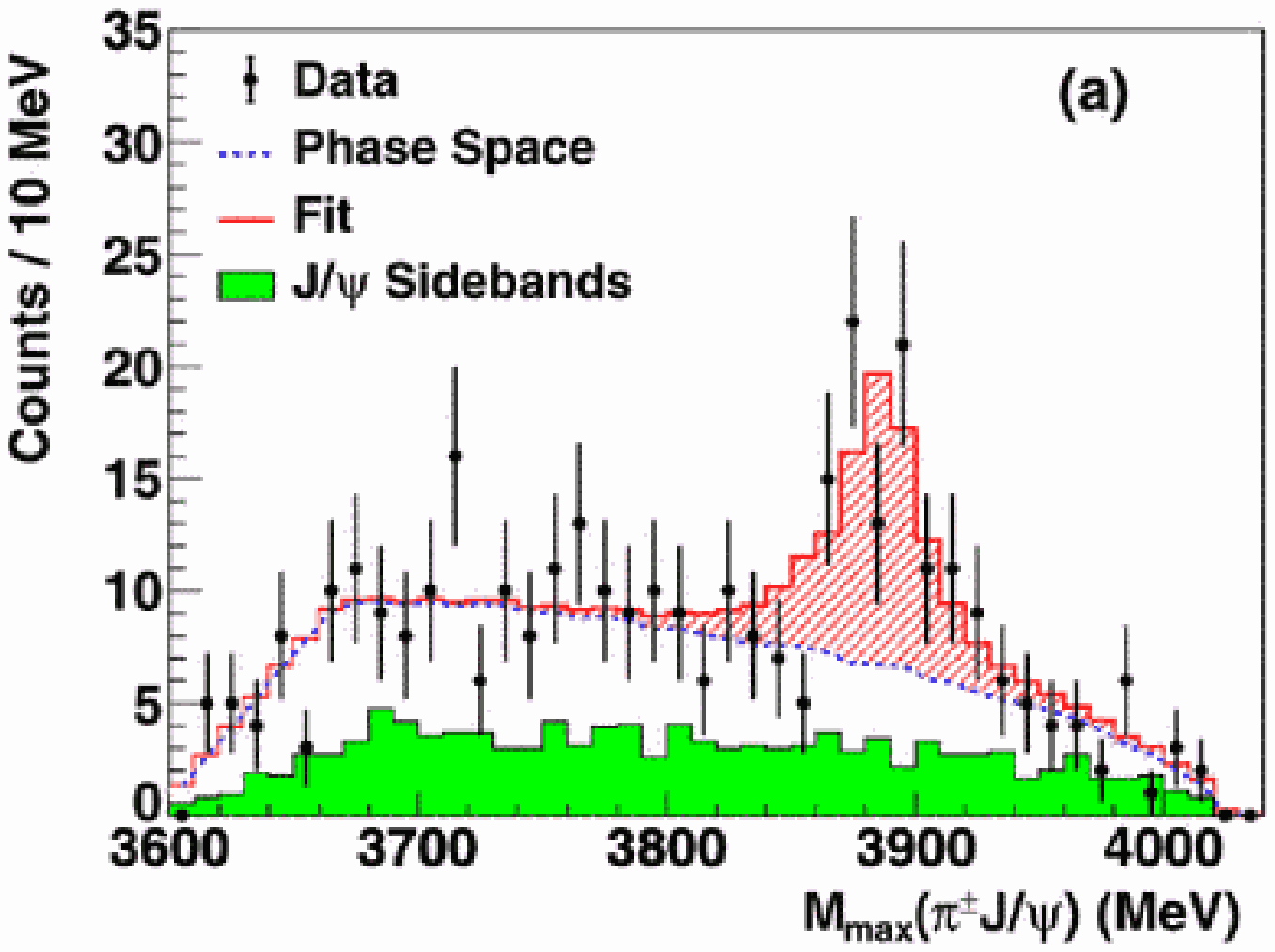}}
\subfigure[]{\label{fig:zcb}\includegraphics[width=0.3\textwidth,height=0.27\textwidth]{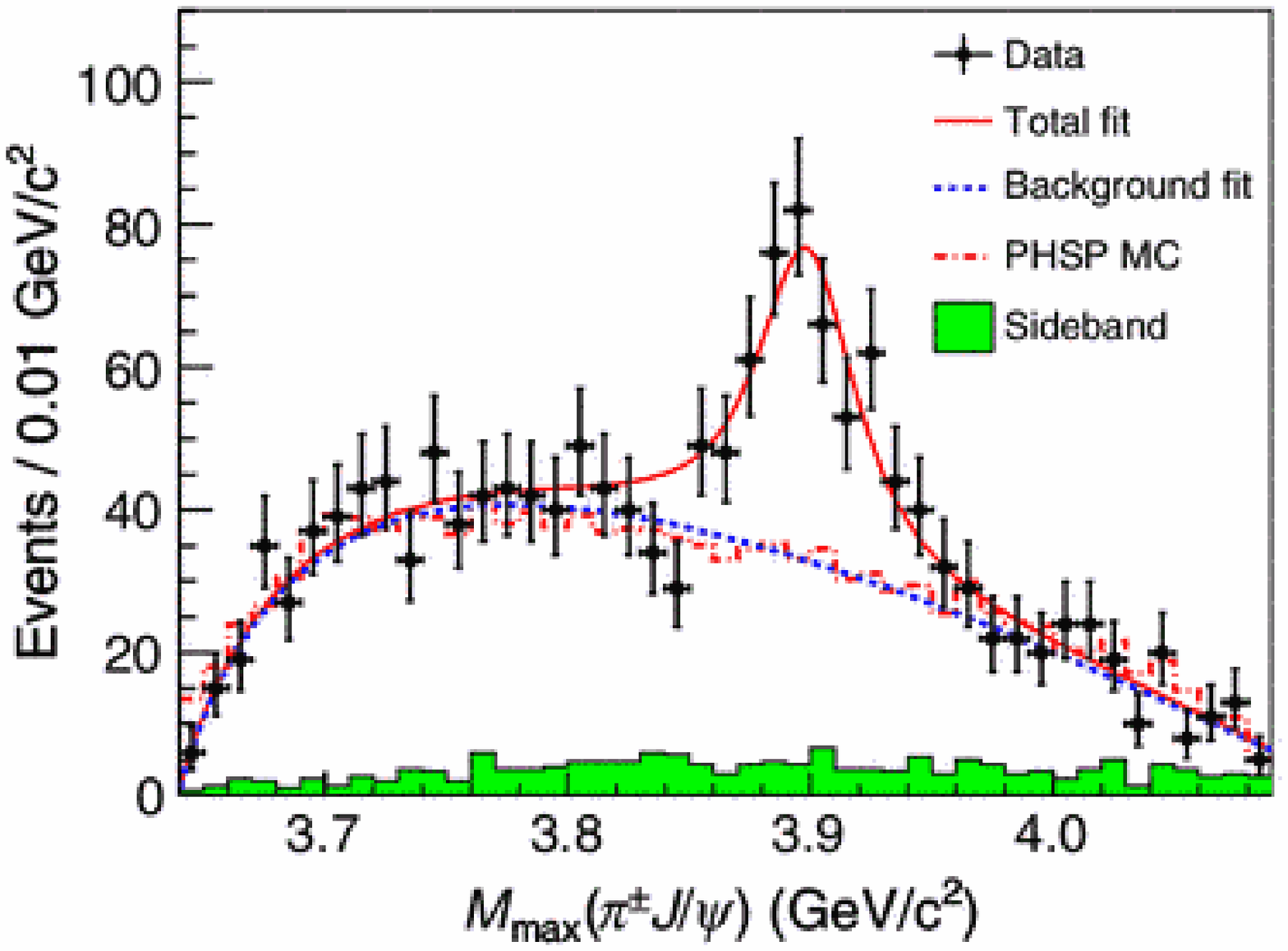}}
\subfigure[]{\label{fig:zcc}\includegraphics[width=0.3\textwidth,height=0.27\textwidth]{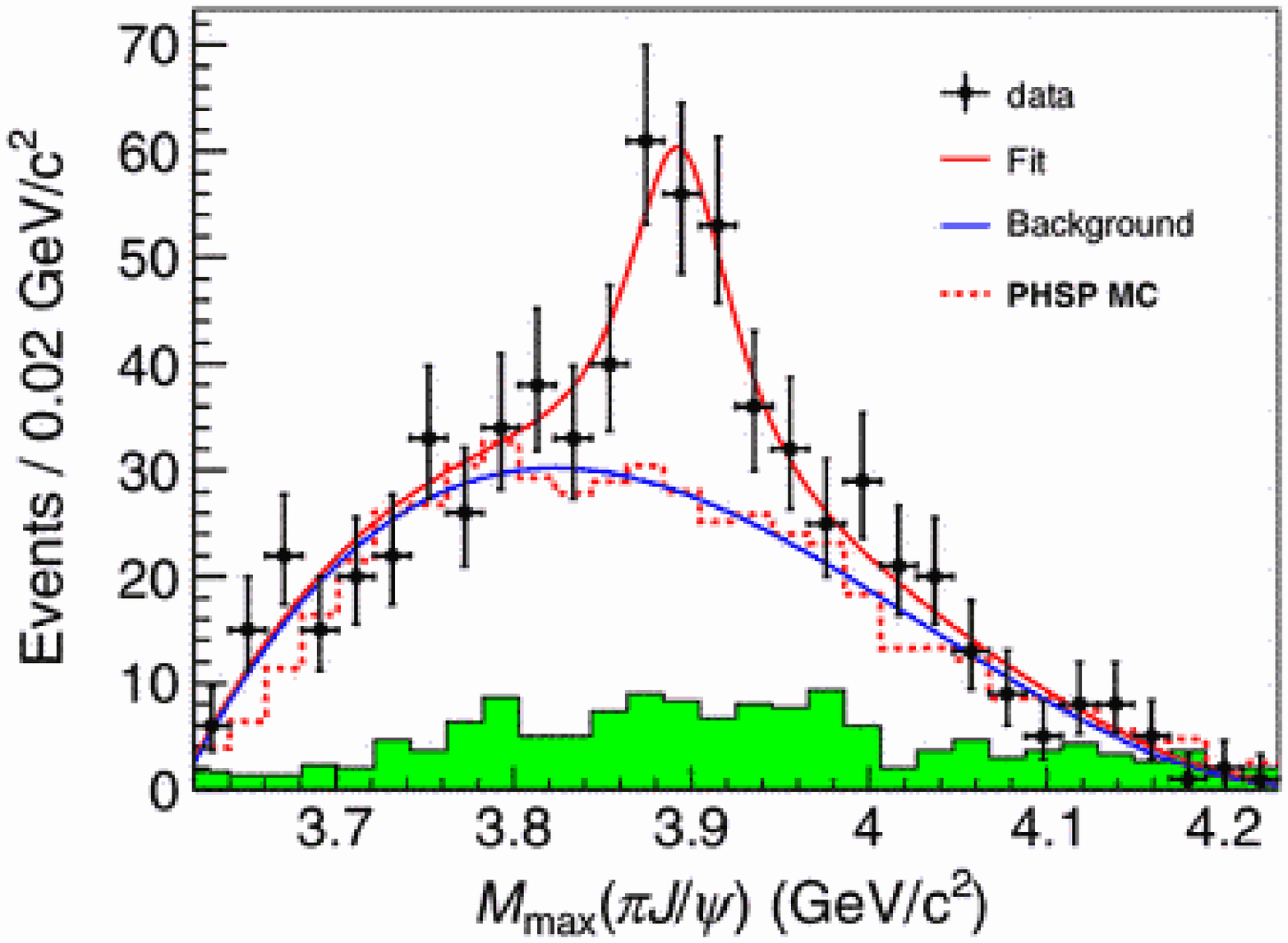}}
\caption{Shapes of $Z_c(3900)$, $M_{max}(\pi^\pm J/\psi)$ from (a) CLEO-c data, (b) BESIII; (c) Belle.}
\label{fig:zc}
\end{figure}

A charged Z state named $Z_c(3900)$ is observed by BESIII~\cite{PhysRevLett.110.252001}
and Belle~\cite{PhysRevLett.110.252002} via $\pi^\pm J/\psi$, and confirmed with CLEO-c's
data~\cite{PhysLett.B727.366}. The measured mass, width, number of events and significance
of the three experiments are shown in Table.~\ref{tab:zc}, the shapes of the
$M_{max}(\pi^\pm J/\psi)$ are shown in Fig.~\ref{fig:zc}. Analysis shows $Z_c(3900)$ is
strongly coupled to $c\bar{c}$ as well as has electric charge, that indicates it at least
a $4-$quarks state. Many interpretations are proposed, such as $DD^*$ module, tetra-quark
state, Cusp, and threshold effect, \etal, but none of them is a satisfied explanation yet.

\begin{table}[ht!]
\begin{center}
\begin{tabular}{c|c|c|c|c}
Exp. & $M(\mathrm(MeV)$ & $\Gamma (\mathrm{MeV})$ & Num. of Evt. & significance 
\\ \hline
BESIII & $3899.0 \pm 3.6 \pm 4.9$ & $46 \pm 10 \pm 20$ & $307 \pm 48$ & $>8 \sigma$ 
\\
Belle & $3894.5 \pm 6.6 \pm 4.5$ & $63 \pm 24 \pm 26$ & $159 \pm 49$ & $>5.2 \sigma$ 
\\
CLEO-c data & $3885 \pm 5 \pm 1$ & $34 \pm 12 \pm 4$ & $81 \pm 20$ & $6.1 \sigma$  
\\ \hline 
\end{tabular}
\caption{Mass, width, number of events and significance of the $Z_c(3900)$ from three experiments.}
\label{tab:zc}
\end{center}
\end{table}

\begin{figure}[ht!]
\centering
\subfigure[]{\label{fig:zc2a}\includegraphics[width=0.3\textwidth,height=0.27\textwidth]{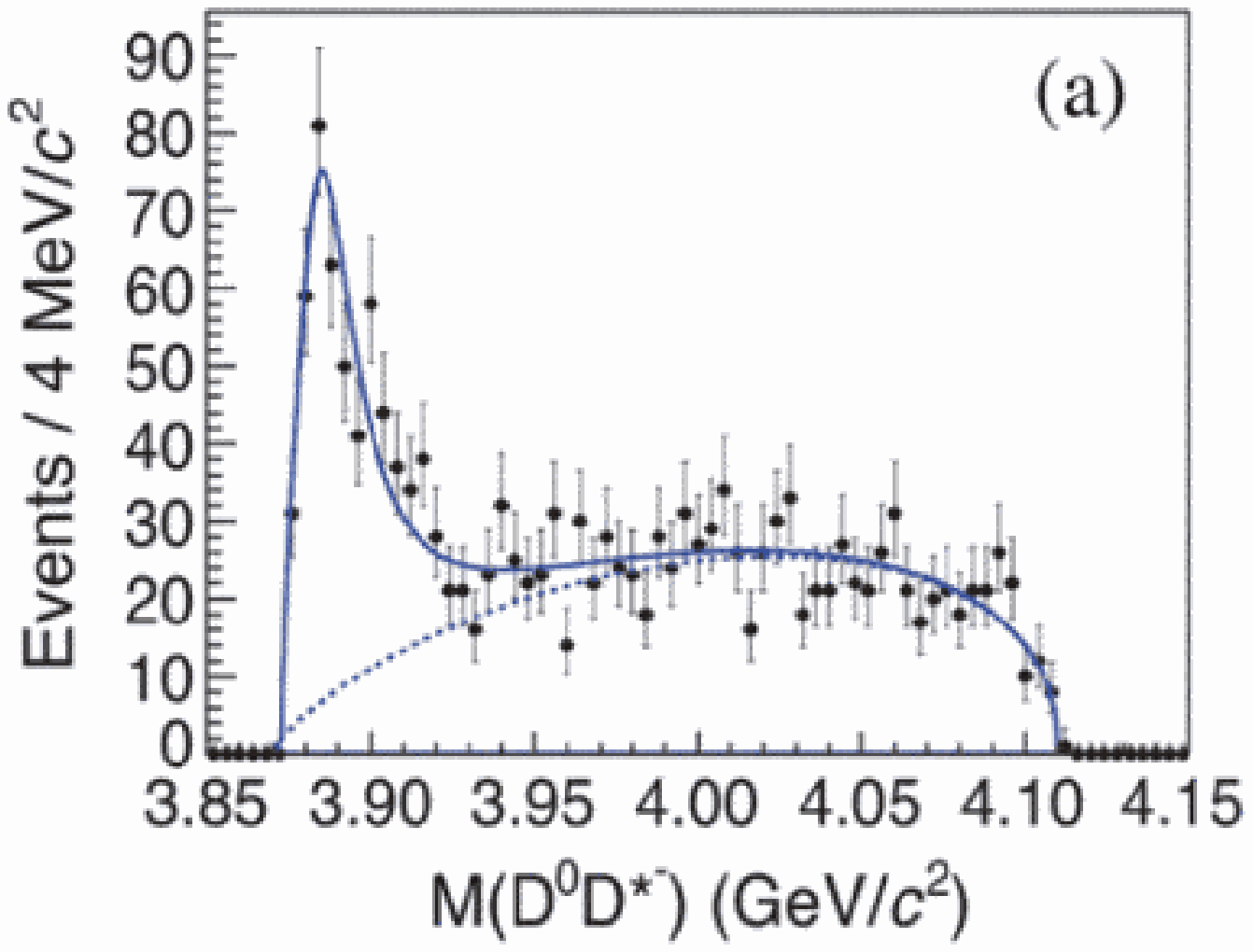}}
\subfigure[]{\label{fig:zc2b}\includegraphics[width=0.3\textwidth,height=0.27\textwidth]{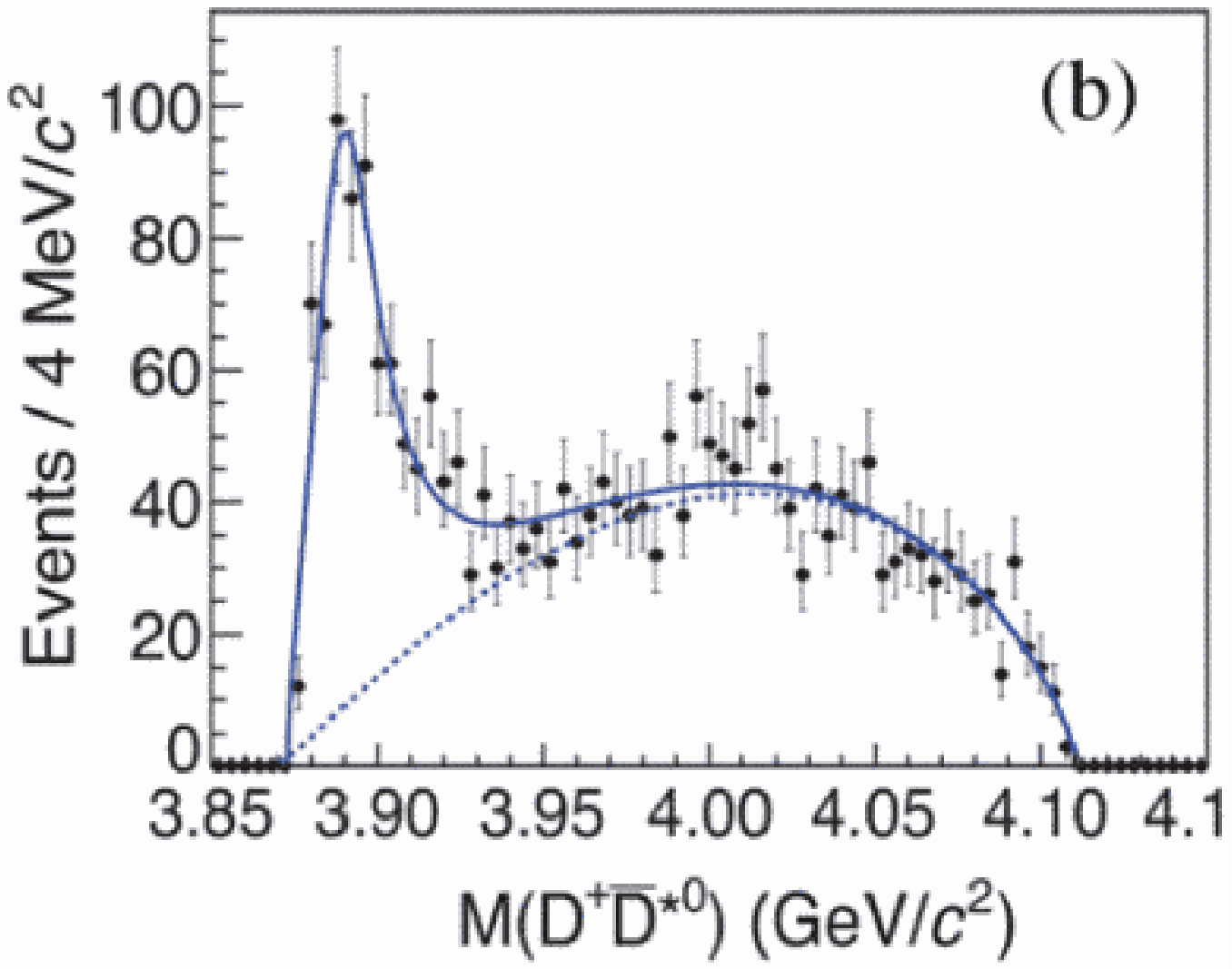}}
\subfigure[]{\label{fig:zc2c}\includegraphics[width=0.3\textwidth,height=0.27\textwidth]{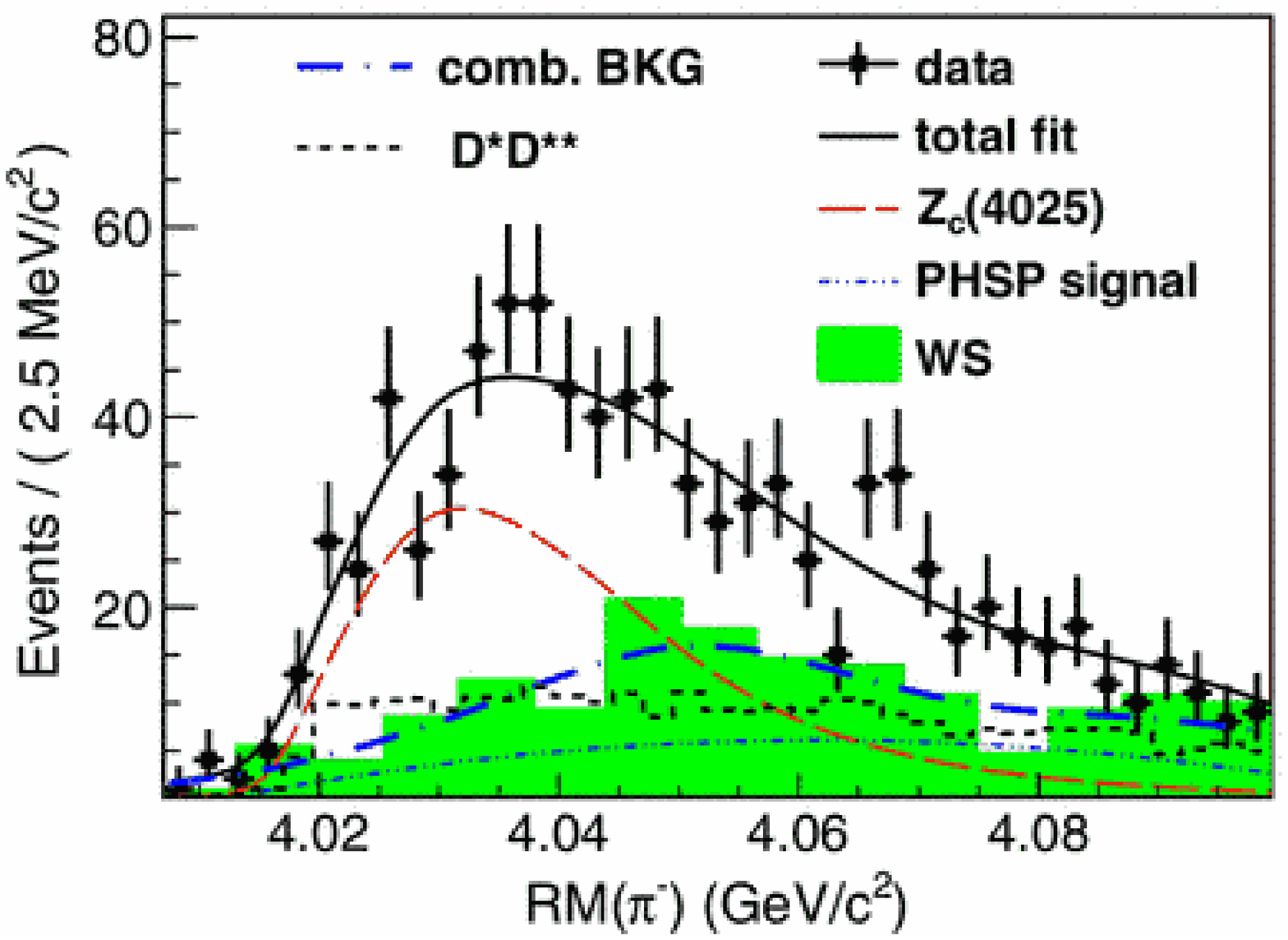}} \\
\subfigure[]{\label{fig:zc2d}\includegraphics[width=0.3\textwidth,height=0.27\textwidth]{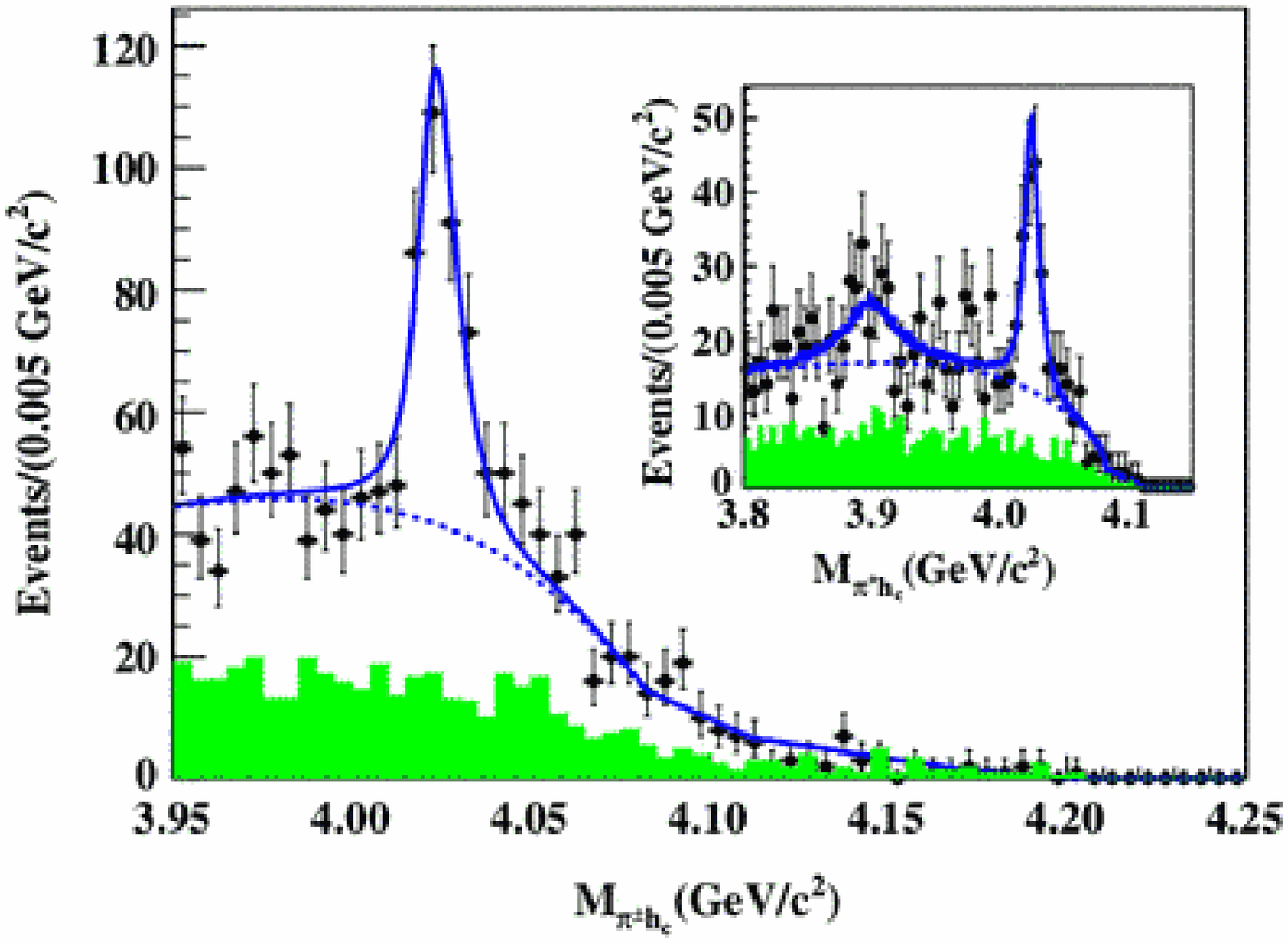}}
\subfigure[]{\label{fig:zc2e}\includegraphics[width=0.3\textwidth,height=0.27\textwidth]{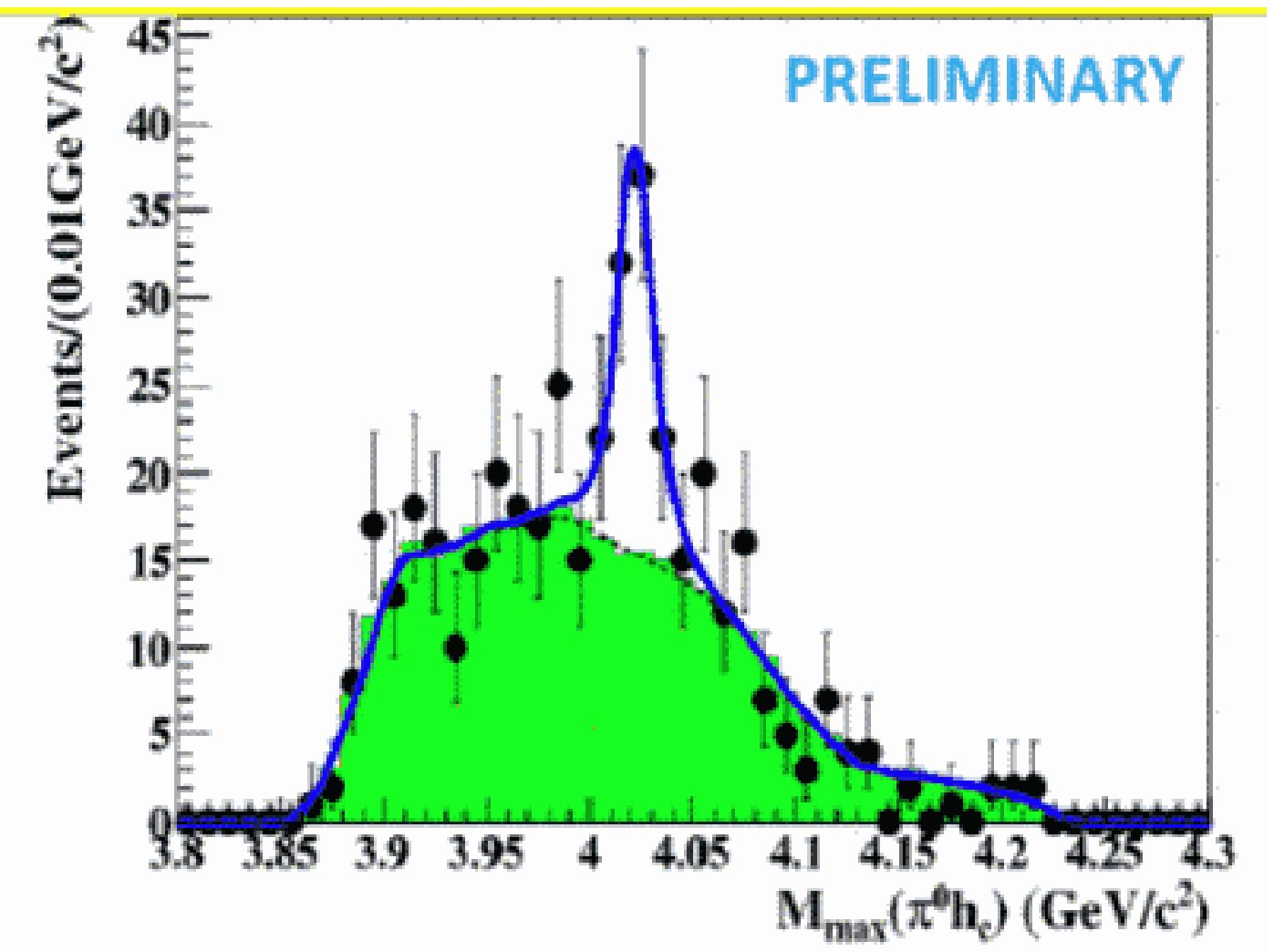}}
\caption{Shapes of (a)(b) $Z_c(3885)$; (c) $Z_c(4025)$; (d) $Z_c^\pm(4020)$, (e) $Z_c^0(4020)$.}
\label{fig:zc2}
\end{figure}

Other $Z_c$ production modes or partners are searched at BESIII with $\pi^\pm
(D\bar{D}^*)^\mp$~\cite{PhysRevLett.112.022001}, $\pi^\pm
(D^*\bar{D}^*)^\mp$~\cite{PhysRevLett.112.132001}, $\pi^+ \pi^-
h_c$~\cite{PhysRevLett.111.242001}, and $\pi^0 \pi^0 h_c$. 
The so called $Z_c(3885)$, $Z_c(4025)$, charged $Z_c(4020)$ and neutral $Z_c(4020)$ are
found; their mass and width are listed in Table~\ref{tab:zc2}, and their shapes are shown
in Fig.~\ref{fig:zc2}. Belle observed $Z_c(4030)$ via $B\to \psi' K \pi^-$ ($K=K^0_s
or K^+$)~\cite{PhysRevLett.100.142001}, and it's confirmed by LHCb via $B^0 \to \psi' \pi
K^+$~\cite{PhysRevLett.112.222002}, as well as its $J^P$ is determined as $1^+$.  

\begin{table}[ht!]
\begin{center}
\begin{tabular}{c|c|c}
states & $M(\mathrm(MeV)$ & $\Gamma (\mathrm{MeV})$ 
\\ \hline
$Z_c(3885)$ & $3883.9 \pm 1.5 \pm 4.2$ & $24.8 \pm 3.3 \pm 11.0$ 
\\
$Z_c(4025)$ & $4026.3 \pm 2.6 \pm 3.7$ & $24.8 \pm 5.6 \pm 7.7$ 
\\
$Z_c^\pm(4020)$ & $4022.9 \pm 0.8 \pm 2.7$ & $7.9 \pm 2.7 \pm 2.6$ 
\\
$Z_c^0(4020)$ & $4023.6 \pm 2.2 \pm 3.9$ & fixed to $\Gamma(Z_c^\pm(4020))$
\\ \hline 
\end{tabular}
\caption{Mass and width of other $Z_c$ states.}
\label{tab:zc2}
\end{center}
\end{table}

In order to search $Z_{cs}$, Belle updated its previous $K^+ K^- J/\psi$ measurement to a
Dalitz Plot analysis~\cite{PhysRev.D89.072015}, no evident structure is found in $K^\pm
J/\psi$ mass distribution under current statistics.

\begin{figure}[ht!]
\centering
\includegraphics[width=0.8\textwidth]{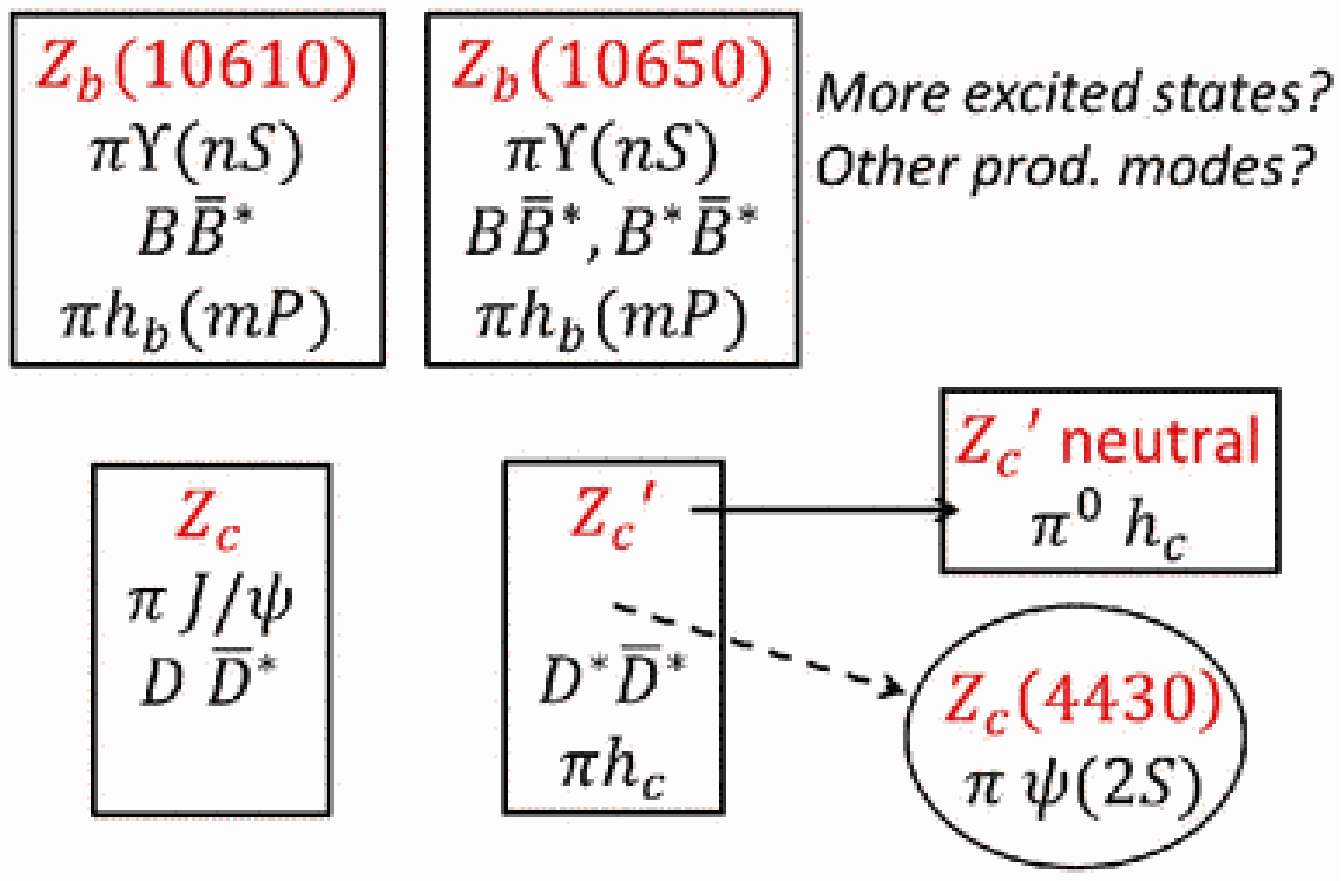}
\caption{Comparison of production modes of $Z_c$ and $Z_b$ states.}
\label{fig:zb}
\end{figure}

Belle has observed $Z_b(10610)$ and $Z_b(10650)$ in $\pi^+ \pi^-
\Upsilon(nS)$~\cite{PhysRevLett.108.122001, arxiv:1105.4583}, $\pi^+ \pi^-
h_b(mP)$~\cite{PhysRevLett.108.032001} and $[B\bar{B}^*]^\pm \pi^\mp$~\cite{arxiv:1209.6450}
final states. Their possible relations to $Z_c$ states are displayed in Fig.~\ref{fig:zb},
here we assume $Z_c(3900)$ and $Z_c(3885)$ are a same state $Z_c$, and $Z_c(4020)$ and
$Z_c(4025)$ are a same state $Z'_c$. From this plot, more excited states or production
mode of Z particles are expected.

\section{summary}
Leptonic machines produce exotics. These particles are unexpected, weird and strange,
while also tantalizing, charming and interesting, and only very limited topics of these
exotics is covered in this proceeding. Some abnormal structures near $p\bar{p}$ threshold
are observed with similar masses, the significances of these signals are large. However,
there are many questions need answer. Are they from a same source? Are they baryonia,
glueball or hybrid? For some cases, PWA is needed to determine the spin and parity. Search
via more decay modes will help to reveal the veil. Lots of and very fast progresses in XYZ
studies with $e^+ e^-$ experiments in recent years, only some new measurements are
reviewed in this proceeding. It includes: a new production mode of $X(3872)$; some new
information on the Y states from Babar, Belle and BESIII. Till now, most of the Y states
are charmonium related, more decays modes of them worthy checking. There are also
contradictive conclusions from different experiment collaborations due to limited
statistics. $Z_c(3900)$, the first confirmed at least four-quarks state, and its partners
has been presented also. Their nature are still unknown. A systematic study of its quantum
number, decay modes, excited states, comparison with their beauty-counterparts is crucial
to achieve a satisfactory explanation. With more upcoming data of the leptonic machines,
an exciting future is expected.

\Acknowledgements 
I am grateful to the FPCP2014 committee for the organization of this nice conference.

 
\end{document}